\documentclass{aa}  
\usepackage{graphicx}
\usepackage[dvipsnames]{xcolor}
\usepackage{txfonts}
\usepackage{color}
\usepackage{soul}
\begin{document}

   \title{VEGAS: a VST Early-type GAlaxy Survey}

   \subtitle{V.~IC~1459 group: Mass assembly history in low density environments}

   \author{Enrichetta Iodice\inst{1,2}
          \and
          Marilena Spavone\inst{1}
          \and
          Arianna Cattapan\inst{1}
          \and
          Elena Bannikova\inst{3,4}
          \and
          Duncan A. Forbes\inst{5}
          \and
          Roberto Rampazzo\inst{6}
          \and 
          Stefano Ciroi\inst{7,8}
          \and 
          Enrico Maria Corsini\inst{7,8}
          \and
          Giuseppe D'Ago\inst{9}
          \and 
          Tom Oosterloo\inst{10,11}
          \and
          Pietro Schipani\inst{1}
          \and 
          Massimo Capaccioli\inst{12}
          }

   \institute{INAF $-$ Astronomical Observatory of Capodimonte, Salita Moiariello 16, I-80131, Naples, Italy\\ % Cattapan, Spavone, Iodice, Schipani
              \email{enrichetta.iodice@inaf.it}
              \and
              European Southern Observatory, Karl-Schwarzschild-Strasse 2, D-85748 Garching bei Muenchen, Germany
              %Iodice
              \and Institute of Radio Astronomy of National Academy of Sciences of Ukraine, Mystetstv~4, UA-61022 Kharkiv, Ukraine %Bannikova
              \and
              V. N. Karazin Kharkiv National University, Svobody Sq.~4, UA-61022 Kharkiv, Ukraine %Bannikova
              \and
              Centre for Astrophysics and Supercomputing, Swinburne University of Technology, Hawthorn, Victoria 3122, Australia %Forbes
              \and
              INAF $-$ Astronomical Observatory of Padova, Via dell’Osservatorio 8 , I-36012, Asiago (VI), Italy %Rampazzo
              \and
              Department of Physics and Astronomy “G. Galilei”, University of Padova, Vicolo dell’Osservatorio 3, I-35122, Padova, Italy %Ciroi, Corsini
              \and
              INAF $-$ Osservatorio Astronomico di Padova, Vicolo dell'Osservatorio 5, I-35122 Padova, Italy %Corsini
              \and
              Instituto de Astrofísica, Facultad de Física, Pontificia Universidad Católica de Chile, Av. Vicu\~{n}a Mackenna 4860, 7820436 Macul, Santiago, Chile %D'Ago
              \and
              ASTRON, Netherlands Institute for Radio Astronomy, Postbus 2, 7990 AA Dwingeloo, The Netherlands %Oosterloo
              \and
              Kapteyn Astronomical Institute, University of Groningen, Postbus 800, 9700 AV Groningen, The Netherlands %Oosterloo
              \and
              University of Naples Federico II, C.U. Monte Sant'Angelo, via Cinthia, I-80126, Naples, Italy %Capaccioli
             }

   \date{Received ???; accepted ???}

% \abstract{}{}{}{}{} 
% 5 {} token are mandatory
 
  \abstract
   {This paper is based on the multi-band VST Early-type GAlaxy Survey (VEGAS) with the VLT Survey Telescope (VST). We present new deep photometry of the IC~1459 group in {\it g} and {\it r} band.}
   {The main goal of this work is to investigate the photometric properties of {  the IC~1459 group, and to compare our results with those obtained for other galaxy groups studied in VEGAS, in order to provide a first view of the variation of their properties as a function of the  evolution of the system.}}
   {{  For all galaxies in the IC~1459 group, we fit isophotes and extract the azimuthally-averaged surface-brightness profiles, the position angle and ellipticity profiles as a function of the semi-major axis, as well as the average colour profile. In each band, we estimate the total magnitudes, effective radii, mean colour, and total stellar mass for each galaxies in the group. Then we look at the structure of the brightest galaxies and faint features in their outskirts, considering also the intragroup component.}}
   {{  The wide field of view, long integration time, high angular resolution, and arcsec-level seeing of OmegaCAM@VST allow us to map the light distribution of IC~1459 down to a surface brightness level of $29.26$~mag~arcsec$^{-2}$ in {\it g} band and $28.85$~mag~arcsec$^{-2}$ in {\it r} band, and out to $7-10 R_{\rm e}$, and to detect the optical counterpart of HI gas around IC~1459. We also explore in depth three low density environments and provide information to understand how galaxies and groups properties change with the group evolution stage.}}
   {There is a good agreement of our results with predictions of numerical simulations regarding the structural properties of the brightest galaxies of the groups. {  We suggest that 
   the  structure  of  the  outer  envelope  of  the  BCGs  (i.e.  the  signatures of past mergers 
   and tidal interactions), the intra-group light and the HI amount and distribution may be used 
   as indicators of the different evolutionary stage and mass assembly in galaxy groups.}}

               
\titlerunning{IC~1459 group: Mass assembly history in low density environments}
\authorrunning{E. Iodice et al. }
\maketitle
%
%-------------------------------------------------------------------

\section{Introduction}\label{sec:intro}

The hierarchical accretion scenario is one of the main products of the $\Lambda$CDM model, with structures forming as the result of merging of smaller elements \citep{deLucia2006}. 
In this framework, at a first epoch the galaxies lie along cosmic web filaments, then due to gravity they fall into small galaxy groups, and these low density environments merge into galaxy clusters \citep{Rudick2009,Mihos2015}. 
{  These small-size environments contain more than $\sim60\%$ of the galaxies in the Universe, and here galaxies spend a meaningful part of their evolutionary life \citep{Miles2004}. 
According to \citet{Bower2004}, in the local universe, the difference between 
groups and cluster of galaxies is based on the virial mass: a group of galaxies it is 
in the range $10^{13} - 10^{14} M_{\odot}$.

During the infall of galaxy groups to form a cluster, the material stripped from galaxy outskirts builds up the intra-group light (IGL) and intra-cluster light \citep[ICL,][]{Fujita2004,Willman2004,Contini2014,DeMaio2018}.
The IGL is the precursor of the ICL \citep{Mihos2015}.
Since ICL/IGL is the fossil record of all past interactions and mergers,
the constant evolution and growth of the ICL/IGL over time with the infalling of galaxies in the 
potential well of the brightest cluster/group galaxy (BCG/BGG), suggests that the ICL/IGL 
properties should be linked to the evolutionary state of the cluster/group \citep{Mihos2015}. 
In this process, the mass assembly in the BCG/BGG is still 
on-going. The imprint of mass assembly in the BCGs/BGGs resides in the stellar halo. 
This is an extended, diffuse, and very faint ($\mu_g \geq 26 - 27$ mag/arcsec$^2$) 
component made of stars stripped from satellite galaxies, in the form of streams and tidal tails, 
with multiple stellar components and complex kinematics \citep[see][for reviews]{Duc2017,Mihos2017}.
Recent theoretical works  provide a detailed set of simulations to reproduce the 
faint features in the galaxy outskirts at comparable levels of the deep observations
\citep[i.e. 29-33 mag/arcsec$^2$,][]{Pop2018,Mancillas2019}.
They made a census of the various types of low surface brightness (LSB) features and traced their 
evolution. According to \citet{Mancillas2019}, the tidal tails are thick elongated structures that 
emerge from the parent galaxy. Differently, the stellar streams are tiny filamentary features that 
originates from the disruption of low-mass satellites in the galaxy halo. 
Shells have arc-like concentric shapes and, depending on the nature and projection, they can appear 
aligned on the same axis.  Tidal tails and shells in the galaxy outskirts results from  
intermediate and major mergers (mass ratio 7:1 to 3:1), whereas the stellar streams are typical 
signatures of minor-mergers. The survival time is estimated between 0.7 up to 4 Gyr, 
where tidal tails have the shorter life time with respect to shells and streams.

Semi-analytic models combined with simulations give detailed predictions 
about the structure and stellar populations of stellar halos, the ICL/IGL formation and 
the amount of substructures in various kinds of environment \citep{Oser2010,Cooper2013,Cooper2015,Cook2016,Pillepich2018,Monachesi2019}. 
Predictions cited above suggest that the 
BCGs/BGGs have an inner stellar component formed {\it in-situ}, whereas the accreted {\it ex-situ} 
component contains all the accreted material.
The {\it ex-situ} component is made by the relaxed component, which is completely merged with the 
{\it in-situ} component, and by the un-relaxed component, which is the outer stellar envelope. 
Simulations show that in the surface-brightness radial profile of simulated galaxies 
there is evidence of {\it inflection} in the region of the stellar halos, corresponding 
to variations in the ratio between the accreted relaxed and the accreted 
un-relaxed components \citep{Cooper2010,Deason2013,Amorisco2017}. 
This distance from the galaxy centre where the inflection occurs 
is the {\it transition radius} ($R_{\rm tr}$) used to 
characterise stellar halo. Massive galaxies with an high accreted mass fraction have a small 
$R_{\rm tr}$ \citep{Cooper2010,Cooper2013}. The un-relaxed component of the stellar envelope appears as a change in the 
slope at larger radii of the surface brightness profiles.
In this context, the study of the surface brightness profiles of BCG/BGG at the faintest 
levels is potentially one of the main "tools" to quantify the contribution of the accreted mass, 
which becomes particularly efficient when the outer stellar envelope starts 
to be dominant beyond the transition radius
\citep{Iodice2016,Iodice2017b, Spavone2017a, Spavone2018}.}

%-----------------------------------------------------------------------------------------
In the last two decades, a huge enhancement to the study of BCGs/BGGs and on the ICL 
in different type of environments has been given by deep imaging surveys aimed at studying galaxy 
structures out to the regions where the galaxy light merges into the intra-cluster component \citep{Ferrarese2012,vanDokkum2014,Duc2015,Munoz2015,Merritt2016,Mihos2017}.
{  The {\it VST Early-type Galaxy Survey} \citep[VEGAS\footnote{see \url{http://www. na.astro.it/vegas/VEGAS/Welcome.html}}, ][]{Capaccioli2015} has occupied in the last years a pivotal role in this field. VEGAS is a multi-band {\it ugri} imaging survey 
with the VLT Survey Telescope (VST). Taking advantage of the large field-of-view of OmegaCAM@VST, VEGAS data allow to relate the 
galaxy structure with environment, from the dense cluster of galaxies \citep[see][and reference therein]{Iodice2019} to the 
unexplored poor groups of galaxies \citep{Spavone2018,Cattapan2019}. 
With the VEGAS data we are able to map the surface brightness of galaxies down to $\mu_g \sim30$~mag/arcsec$^2$ and out to about 10$R_e$ \citep{Spavone2017a,Iodice2019}. 
The deep photometry allows to trace the mass assembly in galaxies, by estimating the accreted mass 
fraction in the stellar halos, detecting the ICL and the stellar streams in the intra-cluster space, 
and providing results that can be directly compared with the predictions of galaxy formation models \citep{Iodice2016,Iodice2017a,Iodice2017b,Spavone2017a,Spavone2018,Cattapan2019}. 
Recently, in the deep imaging data of VEGAS for the NGC~5846 group of galaxies, 
we were able to detect an ultra diffuse galaxy, with an absolute magnitude of $M_g = -14.2$~mag, corresponding to a stellar mass of $\sim 10^8$~M$_\odot$ \citep{Forbes2019}. }

%-----------------------------------------------------------------------------------------
{  In this work we present a new VEGAS deep mosaic of $1 \times 2$ square degrees of the group of 
galaxies centred on the BGG IC~1459. We use {\it g}, {\it r}, and {\it i} images to analyse the 
structure of the group members, to detect any LSB features in the BGG outskirts and in the 
intra-group space. 
Results are compared with those obtained for other two galaxy groups studied in VEGAS, 
 the NGC~5018 group \citep{Spavone2018} and NGC~1533 triplet \citep{Cattapan2019}, since data have comparable depth and were analysed using the same methods and tools.}

%-----------------------------------------------------------------------------------------
The paper is organised as follow. In Section~\ref{subsec:1459group} we describe the IC~1459 group and its main properties. 
The observing strategy and the data reduction procedure are reported in Section~\ref{sec:oss}. 
In Section~\ref{sec:sp} we present the data analysis and in Section~~\ref{sec:1459groupana} 
we describe the results for the IC~1459 group. 
In Section~\ref{sec:lgg} we compare the results of the three analysed VEGAS groups with theoretical predictions and with previous observational results. In Section~\ref{sec:conc} we draw our conclusions.

\section{The IC~1459 Group}\label{subsec:1459group}

\begin{figure*}
	\centering
	\includegraphics[width=13cm]{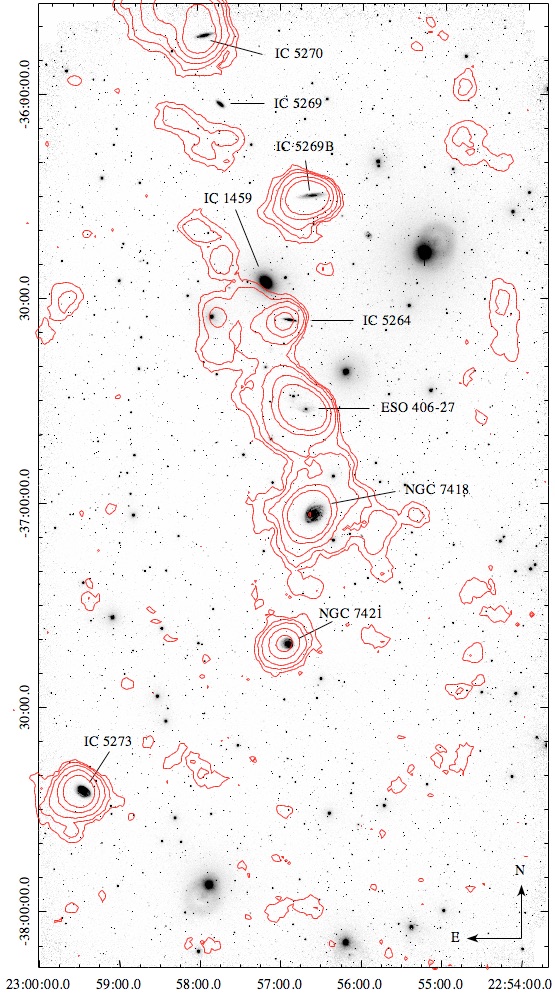}
	\caption{OmegaCAM@VST mosaic in {\it g} band of the IC~1459 group with the HI map from the KAT$-7$ observations (red contours). The image size is about $1^{\circ} \times 2^{\circ}$, and the HI contour levels are $5.5$, $10$, $20$, $50$, and $100 \times 10^{18}$~cm$^{-2}$ \citep[as shown by ][]{Oosterloo2018}. The right ascension and declination (J2000) are given in the horizontal and vertical axis of the field of view, respectively. The north is at the top and east on the left.}
	\label{fig:mosaic}
\end{figure*}

The IC~1459 group (also known as LGG~466) hosts $9$ bright galaxies, of which $7$ are LTGs \citep{Brough2006}. 
IC~1459 is an ETG located in the group projected centre and it is considered the BGG of this group \citep{Saponara2018}. 
{  By adopting the virial radius given by \citet{Brough2006}, $r_{200} = 0.21$ Mpc, we derived 
the virial mass for the group  $M_{200} \simeq 3.7 \times\ 10^{13} M_{\odot}$.
In this work, we studied the brightest galaxies in the range of magnitude $-23 \leq M_g \leq 19.6$~mag}.
{  IC~1459 is one of the two ETGs of the group. It is the most massive, $\mathcal{M}^{*}_{\rm tot} = 1.0 \times 10^{12}$~$\mathcal{M}_{\odot}$, and luminous, $L_{{\rm tot},g} = 1.77 \times 10^{11}$~$L_{\odot}$, galaxy.}
Table~\ref{tab:mw-prop} lists the main properties of the galaxies in the IC~1459 group and Figure~\ref{fig:mosaic} shows 
the OmegaCAM@VST mosaic in {\it g} band of the group. We consider each of the group members at the same distance of 
IC~1459, $D=28.70$~Mpc, based on the HI data \citep{Brough2006,Serra2015,Saponara2018,Oosterloo2018}.

%---------------------------------------------- Table 1------------------------------------
\begin{table*}
	\centering
	\caption{Basic properties of the galaxies in the IC~1459 group.}
	\label{tab:mw-prop}
	\begin{tabular}{lcccccccccccc}
		\hline\hline
		Galaxy&Morphological Type&R.A. &Decl. & Helio.radial velocity\\
		       &                  & (J2000) & (J2000) & km~s$^{-1}$\\
		\hline
		IC~1459& E3-4&$22^{\rm h}57^{\rm m}10^{\rm s}.61$&$-36^{\rm \circ}27\arcmin 44\farcs0$&$1802$\\
		IC~5269&SAB0(rs):&$22^{\rm h}57^{\rm m}43^{\rm s}.66$&$-36^{\rm \circ}01\arcmin 34\farcs4$&$1967$\\
		IC~5269B	&SB(rs)cd:&$22^{\rm h}56^{\rm m}36^{\rm s}.72$&$-36^{\rm \circ}14\arcmin 59\farcs2$&$1667$\\
		IC~5270&SB(r)cd: (e)&$22^{\rm h}57^{\rm m}54^{\rm s}.94$&$-35^{\rm \circ}51\arcmin 29\farcs0$&$1983$\\
		IC~5264&Sab pec (e)n&$22^{\rm h}56^{\rm m}53^{\rm s}.04$&$-36^{\rm \circ}33\arcmin 15\farcs0$&$1934$\\
		ESO~406-27&SA(rs)d:&$22^{\rm h}56^{\rm m}41^{\rm s}.25$&$-36^{\rm \circ}46\arcmin 21\farcs8$&$2102$\\
		NGC~7418&SAB(rs)cd&$22^{\rm h}56^{\rm m}36^{\rm s}.16$&$-37^{\rm \circ}01\arcmin 48\farcs3$&$1450$\\
		NGC~7421&SB(rs)bc&$22^{\rm h}56^{\rm m}54^{\rm s}.33$&$-37^{\rm \circ}20\arcmin 50\farcs1$&$1792$\\
		IC~5273&SB(rs)cd:&$22^{\rm h}59^{\rm m}26^{\rm s}.70$&$-37^{\rm \circ}42\arcmin 10\farcs4$&$1293$\\
		\hline\hline
	\end{tabular}
\tablefoot{Morphological classifications are from RC3, the Third Reference Catalogue of Bright Galaxies \citep{deVaucouleurs1991}. Coordinates and radial velocities are from NED, the NASA/IPAC Extragalactic Database\footnote{\url{https://ned.ipac.caltech.edu}}.}
\end{table*}
%-------------------------- end Table 1---------------------------------------------------------

{  
In the last decades, the entire group centred on IC~1459 was well studied in a wide wavelength range. 
\citet{Osmond2004} found X-rays emission from a diffuse intragroup medium. 
\citet{Kilborn2009}, by studying the  HI content of the group, pointed out that the gas-rich spirals have 
typical  HI masses, which suggests that the gas removal mechanisms are not yet activated. Therefore,
according to the subsequent analysis by \citet{Serra2015}, the HI distribution seems to be consistent with 
a relatively early stage of group assembly. 
In contrast, the brightest group member IC~1459 shows clear signs of accretion and/or merging events. 
First study of the IC~1459 was provided by \citet{Malin1985}, using photographic plates, which 
pointed out the disturbed morphology in the outskirts, in the form of spiral-like features.
The morphology of IC~1459 appears quite disturbed by the presence of a dust lane in the centre \citep{Forbes1994} and
by shells, plumes, and faint features in the galaxy outskirts \citep{Forbes1995}.
The galaxy hosts an active galactic nucleus with two symmetric radio jets \citep{Tingay2015}.
Stellar kinematics revealed the existence of a fast counter-rotating stellar core that might result from 
the accretion of counter-rotating cold gas streams in early times \citep{Franx1988,Prichard2019}.}

%%---------------------------------------------- Figure 1------------------------------------
%---------------------------------------------- end Figure 1------------------------------------

\section{Observations and Data Reduction} \label{sec:oss}

{  The IC~1459 group is one of the {\it VEGAS} target
\citep[P.I. E. Iodice; ][]{Capaccioli2015}.
VEGAS is a multi-band {\it u}, {\it g}, {\it r} and {\it i} imaging survey obtained with the European Southern Observatory (ESO) Very Large Telescope Survey Telescope (VST). 
VST is a 2.6~m wide field optical telescope \citep{Schipani2012}, equipped with OmegaCAM, a $1^{\circ} \times 1^{\circ}$ camera with a resolution of $0.21$~arcsec~pixel$^{-1}$. 
The data we present were obtained in visitor mode (run IDs: 097.B-0806(B), 098.B-0208(A) and 
0100.B-0168(A)), in dark time. The total integration times and the average FWHM in each band are given in Table~\ref{tab:obs-log}. 
A detailed description of the data reduction, using the dedicated pipelines developed to process OmegaCam 
observations ({\it VST-Tube} and {\it AstroWISE} is provided by \citet{McFarland2013,Grado2012,Capaccioli2015,Spavone2017b,Venhola2018}.

For the IC~1459 group, we have obtained a mosaic of about $1^{\circ} \times 2^{\circ}$, it is shown in Fig.~\ref{fig:mosaic}.
Data were acquired with the {\it step dither} observing strategy, consisting of a cycle 
of short exposures ($\sim$~$150$~sec) on the science target and on an adjacent field (close in space and time) to the 
science frame. This strategy was adopted for other VEGAS targets \citep[as for the NGC~5018 group,][]{Spavone2018} 
and for the Fornax Deep Survey \citep[FDS][]{Iodice2016,Venhola2018} and it guarantees a very accurate estimate 
of the sky background, since an average sky frame was derived for each observing night and then 
subtracted from each science frame. 
With the total exposure times adopted for the observations of IC~1459 group, 
the obtained surface brightness depths for a point source at $5 \sigma$ over an area of 
FWHM=1.26 arcsec are $\mu_g=27.3$~mag, $\mu_r=28.9$~mag and $\mu_i=26.2$~mag 
in the $g$, $r$ and $i$ band respectively. 

By adopting the same method described in \citet{Iodice2016}, 
on the sky-subtracted and stacked images, we estimated any residual fluctuations\footnote{The “residual fluctuations” 
in the sky-subtracted images are the deviations from the background in the science frame with respect to the average 
sky frame obtained by the  empty fields close to the target. Therefore, by estimating them, we obtain an estimate on the 
accuracy of the sky-subtraction step.} 
and the limiting radius from the galaxy centre where the galaxy's light blends into the background.
In short, for each galaxy of the sample and in each band, we extracted the 
azimuthally-averaged intensity profile (using the IRAF task ELLIPSE) on the sky-subtracted mosaic, 
after masking all the bright sources (galaxies and stars) and background objects, and from it 
we estimated the outermost radius, from the centre of the galaxy, where counts are consistent with
the average background level. 
Such a value is the residual by the subtraction of the sky frame,  thus it is very close to zero.
The limiting radius sets the surface brightness limit of the VST light profiles and gives an
estimate on the accuracy of the sky subtraction. 
The fluctuations of the background level are on average in the range $0.6-1$ counts in the {\it g} and {\it r} bands, 
respectively and close to zero counts in the {\it i} band. 
The RMS in the background fluctuations, which quantify how good is the sky subtraction, 
is in the range $0.04-0.06$ counts. RMS and the uncertainties on the photometric calibration ($\sim 0.003 - 0.006$~mag)
are taken into account for computing the total uncertainty on the surface brightness 
magnitudes\footnote{The uncertainty in the surface brightness is calculated with the following formula: $err= \sqrt{(2.5/(adu \times \ln(10)))^2 \times ((err_{adu}+err_{sky})^2) +  err_{zp}^2}$, where
$err_{adu}=\sqrt{adu/N-1}$, with N is the  number of pixels used
in the fit, $err_{sky}$ is the rms on the sky background and $err_{zp}$
is the error on the photometric calibration \citep{Capaccioli2015,Seigar2007}.}
For IC~1459 the limiting radius is 20~arcmin ($\sim 167$~kpc) in the $g$ and $r$ bands, and 10~arcmin ($\sim 83$~kpc) 
in the $i$ band. At these radii, we map the surface brightness down to 
$\mu=29 \pm 1$~mag/arcsec$^2$ in the $g$ and $r$ bands.

To fully account for the broadening effect of the seeing on the light distribution of galaxies,
\citet{Capaccioli2015} characterised the point spread function (PSF) for VST by using stars 
on the acquired VST images. They provided the global PSF surface brightness profile, 
which takes into account the scattered light out to a radial distance comparable to that of the galaxies’ 
major axis diameter \citep[see Fig.~B.1 in][]{Capaccioli2015}.
Where needed, a two-dimensional model of the bright stars, close in projection to the galaxy under study, is derived and subtracted off from the image before performing the analysis of the light distribution (see also Sec.~\ref{sec:sp}). 
}

%---------------------------------------------- Table 2------------------------------------
\begin{table}
\centering
	\caption{Observation log.}
	\label{tab:obs-log}
	\begin{tabular}{l c c c c}
		\hline\hline
		Band& RA (J2000)& Decl. (J2000)& T$_{\rm exp}$& FWHM\\
		& & & (sec)& (arcsec)\\
		(1)&(2)&(3)&(4)&(5)\\
		\hline
		{\it g}&$22^{\rm h}56^{\rm m}51^{\rm s}.36$&$-37^{\rm \circ}00\arcmin 50\farcs4$&$14250$&$1.73$\\
		{\it r}&$22^{\rm h}56^{\rm m}51^{\rm s}.36$&$-37^{\rm \circ}00\arcmin 50\farcs4$&$14400$&$0.89$\\
		{\it i}&$22^{\rm h}56^{\rm m}51^{\rm s}.36$&$-37^{\rm \circ}00\arcmin 50\farcs4$&$12750$&$0.94$\\
		\hline\hline
	\end{tabular}
	\tablefoot{{\it Col.~1} - filters in the SDSS band; {\it Col.~2} and {\it Col.~3} - central right ascension and declination of the mosaic; {\it Col.~4} - total exposure time; {\it Col.~5} - median value of the seeing, FWHM, of the combined frames.}
\end{table}
%-------------------------- end Table 2----------------------------------------------------------

\section{Data Analysis}\label{sec:sp}
\begin{figure*}
	\centering
	\includegraphics[width=18cm]{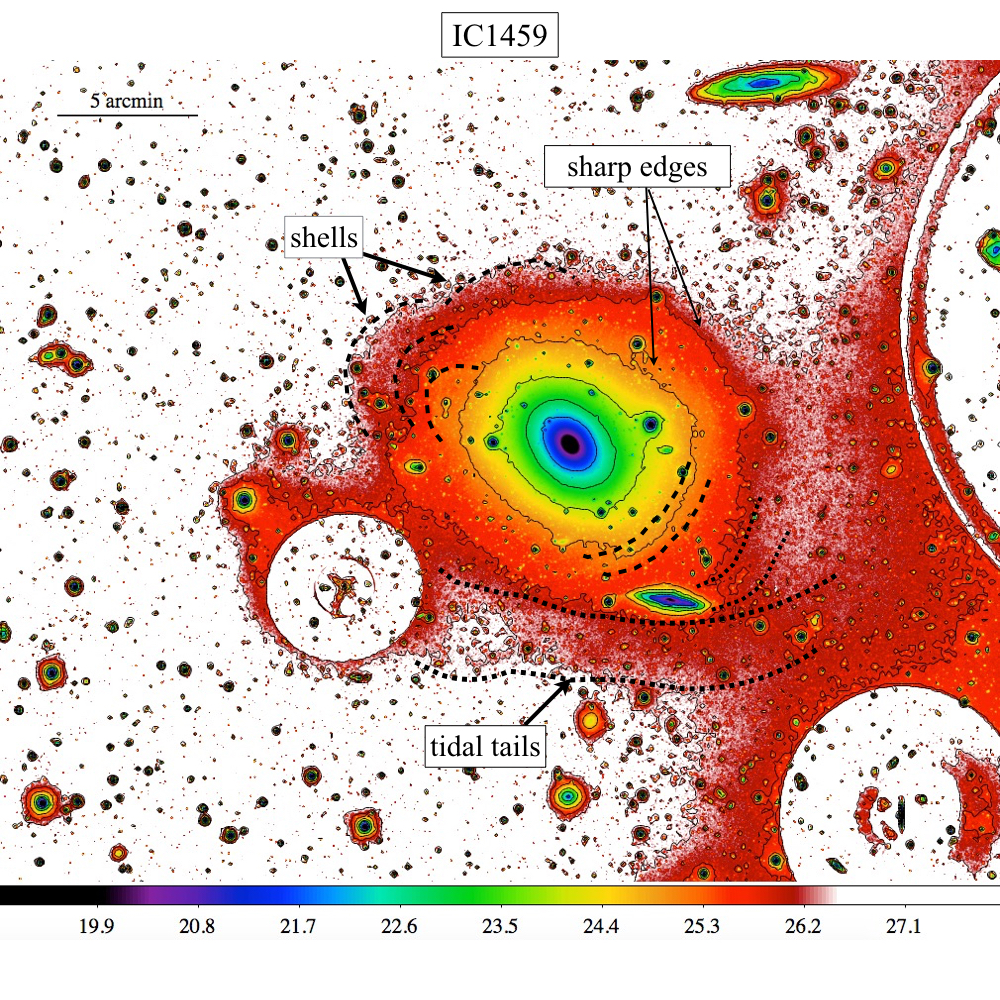}
	\caption{  Enlarged region ($\sim52\arcmin \times 43\arcmin$) of the VST mosaic around IC~1459 in the $g$ band. The image is in surface brightness levels, given in the horizontal colour-bar. The solid contours correspond to the following surface brightness levels: 22, 23, 24, 25, 26.4 mag/arcsec$^2$. The main low surface brightness features detected in the galaxy outskirts are marked on the image: dotted lines and dashed lines delimit the tidal tails and shells, respectively. The prominent sharp edges on the West side are also indicated with arrows (see text for details).}
	\label{fig:1459SB}
\end{figure*}

{  In this section we describe the analysis performed on the new VEGAS data for the IC~1459 group.
We  provide a detailed description of the galaxy structure at the new faintest surface brightness 
levels, focusing on the the outskirts of the BGG IC~1459. In addition, we derive the {\it i)} light profiles 
in all bands and colour distribution for all galaxies of the group; {\it ii)} the contribution to the light of the 
different components in the BGG; {\it iii)} the intra-group light.
Taking advantage of the deep imaging from 
VEGAS, the main science goal of this analysis, is to address the mass assembly history of the IC~1459 group (see Sec.~\ref{sec:1459groupana}).

\subsection{The low surface brightness features in IC~1459 groups}\label{sec:morphology}

In Fig.~\ref{fig:1459SB} we show an enlarged region of the VST mosaic around IC~1459 in the $g$-band surface 
brightness levels\footnote{In this image we modelled and subtracted only the brightest regions of the close 
foreground stars. This prevent the subtraction of any physical faint feature overlapping the halo of the stars.}. 
This is the region of the group where the majority of the faint low surface brightness features 
are detected. All of them are in the outskirt of the brightest group member IC~1459.
The galaxy shows an extended ($\simeq 8$~arcmin) envelope, down to $\mu_g \sim 27$~mag/arcsec$^2$, which appear twisted
with respect to the central brightest region of the galaxy and very irregular in shape.
It is more elongated in the NE-SW direction, where we detected faint ($\mu_g \sim 26.5$~mag/arcsec$^2$) concentric 
shells at distance of $\simeq 5 - 8$~arcmin from the galaxy centre. 
In the same range of radii, on the NW, the envelope has prominent sharp edges at the surface brightness levels of
$\mu_g \sim 24.5 - 26.5$~mag/arcsec$^2$.
An elongated thick structure, of $\simeq 10$~arcmin with $\mu_g \sim 26.5$~mag/arcsec$^2$, 
extends from east to west in the south region of the galaxy. Following \citet[][]{Mancillas2019}, this resembles
a tidal tail. 

The small group member IC~5264 (see Fig.~\ref{fig:mosaic}) in the SW region turns to be completely 
embedded in the IC~1459's envelope. 
We detected an "S-shape" of the disk in IC~1459 with a thick arc-like tail protruding on the West side 
(see Fig.~\ref{fig:1459SB}), of about 3~arcmin and with $\mu_g \sim 26.5$~mag/arcsec$^2$. 

A detailed inspection of the whole VST mosaic does not show any other low surface brightness feature 
(at the imaging depth of the observations) in the intra-group space.
The colour composite images for the other group members, given in Appendix~\ref{apx:B}, show that all galaxies, 
except the faintest S0 IC~5269, have an asymmetric outskirt. In particular,  the spiral galaxy ESO~406-27 show a 
prominent plume emerging from the disk in the NE, at which is associated an over-density of the HI gas (see Fig.~\ref{fig:eso40627sp}). All group members are described in Appendix~\ref{apx:A}.}

\subsection{Isophotal Analysis}\label{subsec:1459ia}
In order to map the light and colour distribution for all group members, we have performed the isophotal analysis 
by using the IRAF task \textsc{ellipse} (the position angle and the ellipticity are free parameters) 
on the final stacked image in each band, {  out to the limiting radius estimated for each object (see Sec.~\ref{sec:oss}). }
The method is the same adopted in the previous studies based on the VEGAS data \citep{Capaccioli2015,Iodice2016,Iodice2017b,Spavone2017b,Spavone2018,Cattapan2019,Iodice2019}.
{  The isophote fit was performed for each galaxy by masking all the bright sources in the field 
(stars and background galaxies). In the case of IC~5264 and IC~5269B, which are close/embedded in the diffuse stellar 
envelope of IC~1459 (see Fig.~\ref{fig:1459SB}), the isophote fit was performed after subtracting the two-dimensional 
model\footnote{The two-dimensional model of IC~1459 is based on the isophote fit and it was derived by using the task 
BMODEL in IRAF.} of IC~1459. }

\begin{figure*}
%	\centering
	\includegraphics[width=9cm]{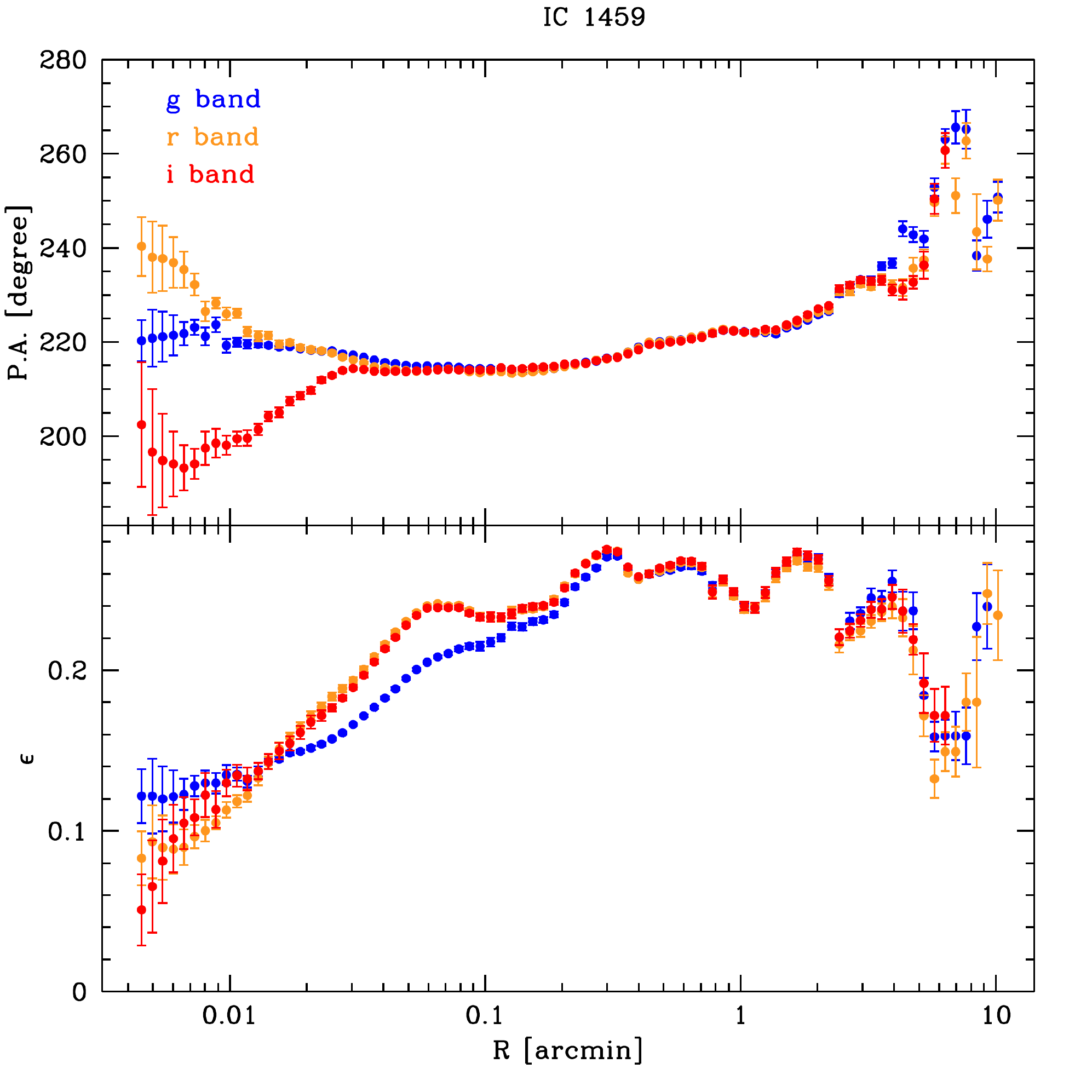}
	\includegraphics[width=9cm]{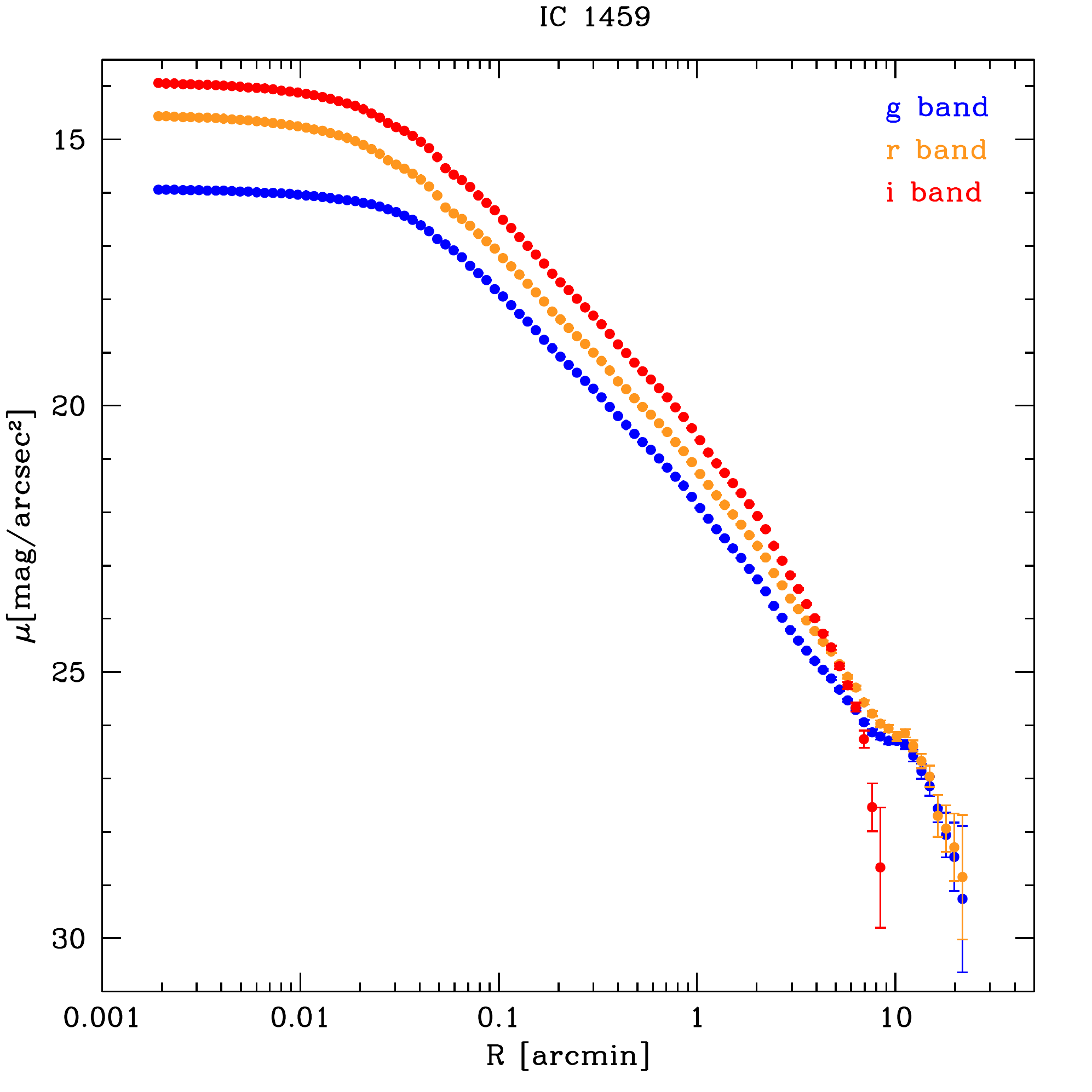}
	\caption{  Results from the isophotal analysis for IC~1459. Ellipticity  and position angle radial profiles are shown in the left lower and top panels, respectively. The azimuthally-averaged and PSF-deconvolved surface-brightness radial profile of IC~1459 are plotted in the right panel.}
	\label{fig:1459profiles}
\end{figure*}

\begin{figure*}
    %centering
	\includegraphics[width=9cm]{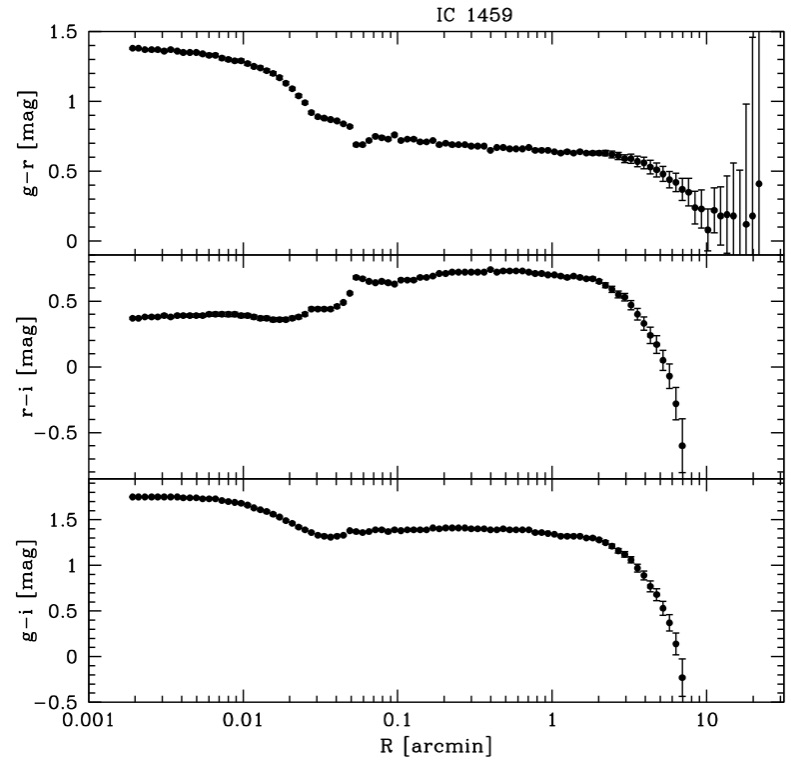}
	\includegraphics[width=9cm]{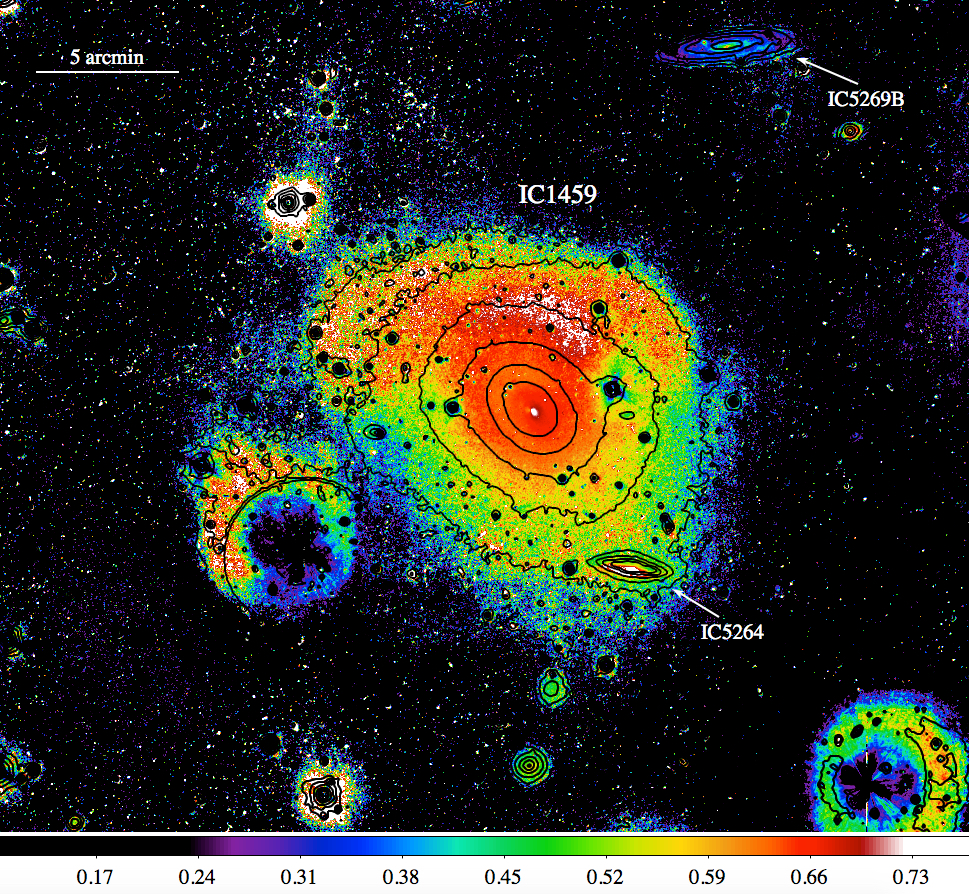}
	\caption{  Left panel: azimuthally-averaged, PSF-deconvolved, and extinction-corrected colour radial profiles of IC~1459. Right panel: two-dimensional colour map centred on IC~1459, the image size is $35.42\farcm7 \times 29.05\farcm7$ with north at the top and east on the left. 
	The horizontal colour bar gives the $g-r$ colour scale. The solid contours correspond to the following surface brightness levels in the $g$ band: 22, 23, 24, 25, 26.4 mag/arcsec$^2$.}
	\label{fig:1459col}
\end{figure*}

From the isophotal analysis, we have derived the azimuthally-averaged surface-brightness, ellipticity ($\epsilon$), 
and position angle (PA) radial profiles in the {\it g} and {\it r} bands, the averaged extinction-corrected colour profiles, and the averaged $g-r$, $r-i$, and $g-i$ colour values. In addition, from the growth curve we have estimated the total magnitude and effective radii in each band.
Finally, using stellar population synthesis models \citep{Ricciardelli2012,Vazdekis2012} with $\log Z/Z_{\odot}=0$ and a Kroupa initial mass function and considering the average colour, we have estimated the total stellar mass by using the M/L ratio in the {\it g} band \citep{Iodice2017b,Spavone2018,Cattapan2019}.
Results are reported in Table~\ref{tab:mtotre} and Table~\ref{tab:colmass}. For the BGG, IC~1459, the 
azimuthally-averaged surface-brightness, ellipticity, and  position angle radial profiles are shown in 
Figure~\ref{fig:1459profiles}. 
Figure~\ref{fig:1459col} shows the azimuthally averaged extinction-corrected colour profile and the two-dimensional $g-r$ colour map centred on IC~1459. 
The surface brightness profiles are PSF-deconvolved \citep[see ][]{Capaccioli2015,Spavone2017b}. 
For the other group galaxies the color-composite image, the azimuthally-averaged surface-brightness radial profiles in {\it g}, {\it r}, and {\it i} band, and azimuthally-averaged extinction-corrected colour profiles are shown in Appendix~\ref{apx:B}.
%

%---------------------------------------------- Table 3------------------------------------
\begin{table*}
%\begin{sidewaystable}
	\centering
	%\scriptsize
	\caption{Total magnitude and effective radius for the galaxies of the IC~1459 group.}
	\label{tab:mtotre}
	\begin{tabular}{lcccccccccccc}
		\hline\hline
		Galaxy&
		$m_{g}$&
		$R_{{\rm e},g}$&
		%\multicolumn{2}{c}{$R_{{\rm e},g}$}&
		%$M_g$&
		$m_{r}$&
		%\multicolumn{2}{c}{$R_{{\rm e},r}$}&
		$R_{{\rm e},r}$&
		%$M_r$&
		$m_{i}$&
		%\multicolumn{2}{c}{$R_{{\rm e},i}$}&
		$R_{{\rm e},i}$\\
		%$M_i$\\
		&
		(mag)&
		(arcsec)&
		%\multicolumn{1}{c}{(arcsec)}&
		%\multicolumn{1}{c}{(kpc)}&
		(mag)&
		%(mag)&
		(arcsec)&
		%\multicolumn{1}{c}{(arcsec)}&
		%\multicolumn{1}{c}{(kpc)}&
		(mag)&
		%(mag)&
		(arcsec)&\\
		%\multicolumn{1}{c}{(arcsec)}&
		%\multicolumn{1}{c}{(kpc)}&
		%(mag)\\
		(1)&
		(2)&
		(3)&
		%\multicolumn{2}{c}{(3)}&
		(4)&
		(5)&
		(6)&
		%\multicolumn{2}{c}{(6)}&
		(7)&\\
		%(8)&
		%(9)&
		%\multicolumn{2}{c}{(9)}&
		%(10)\\
		\hline
		%IC~1459		&$9.28\pm0.07$ &$177.00\pm0.01$&$23.88$&$-23.01$&	$8.71\pm0.07$&$133.60\pm0.81$&$17.94$&$-23.58$&	$8.49\pm0.01$&$50.8\pm0.19$&$7.02$&$-23.80$\\
		%IC~5269B	&$12.69\pm0.02$&$67.42\pm0.81$&$9.04$&$-19.60$&		$12.27\pm0.02$&$68.49\pm0.89$&$9.17$&$-20.02$&	$11.38\pm0.03$&$88.31\pm1.23$&$12.21$&$-20.91$\\
		%IC~5269		&$12.58\pm0.02$&$20.91\pm0.20$&$2.70$&$-19.71$&		$12.01\pm0.02$&$18.59\pm0.33$&$2.43$&$-20.28$&	$11.24\pm0.01$&$18.59\pm0.33$&$2.57$&$-21.05$\\
		%IC~5270		&$12.24\pm0.03$&$47.15\pm1.81$&$6.34$&$-20.05$&		$11.72\pm0.03$&$49.06\pm0.89$&$6.61$&$-20.60$&	$11.04\pm0.05$&$44.16\pm0.79$&$6.11$&$-21.25$\\
		%IC~5264		&$12.67\pm0.03$&$30.21\pm0.44$&$4.05$&$-19.62$&		$11.89\pm0.04$&$34.34\pm0.76$&$4.59$&$-20.40$&	$11.16\pm0.04$&$31.55\pm0.75$&$4.36$&$-21.13$\\
		%ESO~406-27	&$12.86\pm0.06$&$82.58\pm3.18$&$11.06$&$-19.43$&	$12.73\pm0.09$&$75.08\pm5.95$&$10.12$&$-19.56$&	$12.38\pm0.07$&$49.08\pm1.09$&$6.79$&$-19.91$\\
		%NGC~7418	&$10.31\pm0.03$&$161.00\pm2.49$&$21.72$&$-21.98$&	$10.01\pm0.03$&$143.90\pm2.36$&$19.29$&$-22.28$&	$9.94\pm0.01$&$57.48\pm0.37$&$7.95$&$-22.35$\\
		%NGC~7421	&$11.93\pm0.01$&$39.87\pm1.83$&$5.26$&$-20.36$&		$11.53\pm0.01$&$32.27\pm0.07$&$4.32$&$-20.76$&	$10.81\pm0.01$&$31.49\pm0.16$&$4.35$&$-21.48$\\
		%IC~5273		&$10.70\pm0.04$&$99.77\pm2.01$&$13.79$&$-21.59$&	$10.52\pm0.03$&$77.70\pm1.30$&$10.74$&$-21.77$&	$10.23\pm0.02$&$41.11\pm0.38$&$5.68$&$-22.06$\\
		
		IC~1459		&$9.28\pm0.07$ &$177.00\pm0.01$&$8.71\pm0.07$&$133.60\pm0.81$&$8.49\pm0.01$&$50.8\pm0.19$\\
		IC~5269		&$12.58\pm0.02$&$20.91\pm0.20$&$12.01\pm0.02$&$18.59\pm0.33$&$11.24\pm0.01$&$18.59\pm0.33$\\
		IC~5269B	&$12.69\pm0.02$&$67.42\pm0.81$&$12.27\pm0.02$&$68.49\pm0.89$&$11.38\pm0.03$&$88.31\pm1.23$\\
		IC~5270		&$12.24\pm0.03$&$47.15\pm1.81$&$11.72\pm0.03$&$49.06\pm0.89$&$11.04\pm0.05$&$44.16\pm0.79$\\
		IC~5264		&$12.67\pm0.03$&$30.21\pm0.44$&$11.89\pm0.04$&$34.34\pm0.76$&$11.16\pm0.04$&$31.55\pm0.75$\\
		ESO~406-27	&$12.86\pm0.06$&$82.58\pm3.18$&$12.73\pm0.09$&$75.08\pm5.95$&$12.38\pm0.07$&$49.08\pm1.09$\\
		NGC~7418	&$10.31\pm0.03$&$161.00\pm2.49$&$10.01\pm0.03$&$143.90\pm2.36$&$9.94\pm0.01$&$57.48\pm0.37$\\
		NGC~7421	&$11.93\pm0.01$&$39.87\pm1.83$&$11.53\pm0.01$&$32.27\pm0.07$&$10.81\pm0.01$&$31.49\pm0.16$\\
		IC~5273		&$10.70\pm0.04$&$99.77\pm2.01$&$10.52\pm0.03$&$77.70\pm1.30$&$10.23\pm0.02$&$41.11\pm0.38$\\
		\hline\hline
	\end{tabular}
\tablefoot{{\it Col.~1} - Galaxy name. 
{\it Col.~2} and {\it Col.~3} - total magnitude and effective radius in {\it g} band. 
{\it Col.~4} and {\it Col.~5} - total magnitude and effective radius in {\it r} band.
{\it Col.~6} and {\it Col.~7} - total magnitude and effective radius in {\it i} band.}
%\end{sidewaystable}
\end{table*}
%-------------------------- end Table 3---------------------------------------------------------

%---------------------------------------------- Table 3bis------------------------------------
\begin{table*}
	\centering
	\caption{Colours, and total stellar mass for the galaxies of the IC~1459 group.}
	\label{tab:colmass}
	\begin{tabular}{lccccccccc}
		\hline\hline
		Galaxy&
		$A_{g^{\prime}}$&
		$A_{r^{\prime}}$&
		$A_{i^{\prime}}$&
		$g-r$&
		$r-i$&
		$g-i$&
		$(\mathcal{M}/L)_g$&
		$L_{{\rm tot},g}$&
		$\mathcal{M}_{{\rm tot}} ^{*}$\\
		&
		(mag)&
		(mag)&
		(mag)&
		(mag)&
		(mag)&
		(mag)&
		($\mathcal{M}_{\odot} / L_{\odot}$)&
		($\times 10^{10} L_{\odot}$)&
		($\times 10^{10} \mathcal{M}_{\odot}$)\\
		(1)&
		(2)&
		(3)&
		(4)&
		(5)&
		(6)&
		(7)&
		(8)&
		(9)&
		(10)\\
		\hline
		IC~1459		&$0.060$&$0.044$&$0.033$&$0.84\pm0.31$&$0.51\pm0.20$&$1.36\pm0.30$&$5.695$&$17.70$&$100.80$\\
		IC~5269B	&$0.060$&$0.044$&$0.033$&$0.49\pm0.10$&$0.65\pm0.09$&$1.14\pm0.18$&$0.656$&$0.77$&$0.50$\\
		IC~5269		&$0.062$&$0.045$&$0.034$&$0.77\pm0.25$&$0.71\pm0.06$&$1.47\pm0.19$&$3.921$&$0.85$&$3.32$\\
		IC~5270		&$0.052$&$0.038$&$0.029$&$0.72\pm0.09$&$0.65\pm0.12$&$1.38\pm0.15$&$2.335$&$1.16$&$2.71$\\
		IC~5264		&$0.077$&$0.056$&$0.042$&$0.79\pm0.06$&$0.81\pm0.02$&$1.60\pm0.05$&$4.301$&$0.78$&$3.35$\\
		ESO~406-27	&$0.081$&$0.059$&$0.044$&$0.34\pm0.20$&$0.45\pm0.16$&$0.79\pm0.16$&$0.488$&$0.65$&$0.32$\\
		NGC~7418	&$0.060$&$0.044$&$0.033$&$0.66\pm0.29$&$0.44\pm0.57$&$1.09\pm0.74$&$1.585$&$6.86$&$10.87$\\
		NGC~7421	&$0.056$&$0.041$&$0.031$&$0.66\pm0.24$&$0.68\pm0.06$&$1.35\pm0.23$&$1.585$&$1.54$&$2.44$\\
		IC~5273		&$0.047$&$0.034$&$0.026$&$0.54\pm0.15$&$0.51\pm0.28$&$1.05\pm0.40$&$0.701$&$4.78$&$3.36$\\
		\hline\hline
	\end{tabular}
\tablefoot{{\it Col.~1} - Galaxy name. {\it Col.~2}, {\it Col.~3} and {\it Col.~4} - extinction correction in the {\it g}, {\it r}, and {\it i} band from \texttt{NED}.
{\it Col.~5}, {\it Col.~6} and {\it Col.~7} - averaged extinction-corrected $g-r$, $r-i$, and $g-i$ colour value. 
{\it Col.~8} - mass-to-light ration in {\it g} band. 
{\it Col.~9} - total stellar luminosity in {\it g} band. 
{\it Col.~10} - total stellar mass. }
\end{table*}
% -------------------------- end Table

%---------------------------------------------- end Figure 2------------------------------------
%---------------------------------------------- Figure 3------------------------------------
%---------------------------------------------- end Figure 3------------------------------------

\subsection{Fitting of the Light Distribution}\label{subsec:1459ld}

In order to identify the main components dominating the galaxy light in IC~1459, we have fitted the deconvolved 
surface-brightness radial profiles in the {\it g} band. {{  Since our aim is to compare the results of this fit with 
previously published ones, for consistency we use the {\it g} band profile, as done for NGC 5018 \citep{Spavone2018} and 
NGC 1533 \citep{Cattapan2019}.

The main outcome of the fitting is to map the stellar distribution in the outer envelopes of our sample galaxies.
To this aim, according to \citet{Spavone2017b}, we do not use the $\chi^{2}$ statistics in our fitting procedure. The data points corresponding to the central regions of the galaxies, in fact, with their small uncertainties have considerable weight in determining the best-fit solution obtained by minimising the $\chi^{2}$, while the outer regions with bigger errors have no weight. To avoid this problem, we adopt the same approach described by \citet{Seigar2007,Spavone2017b} and performed least-square fits using a Levenberg–Marquardt algorithm, in which the function to be minimised is the rms scatter, defined as $\Delta=\sqrt{ \frac{\sum_{i=1}^{m}
    \delta_{i}^{2}}{m}}$, where $m$ is the number of data points
and $\delta_{i}$ is the {\it i}th residual.}}
{  In order to account for fainter stellar envelope in the most luminous ETGs, 
there is remarkable evidence from the recent deep surveys that the light profiles of these galaxies are not well fitted by a single Sérsic law and at least one additional component is needed to map the outer envelope contribution in the light distribution of the galaxy 
\citep[see e.g. ][]{Seigar2007,Donzelli2011,Arnaboldi2012,Huang2013,Iodice2016,Spavone2017a,Spavone2018}. }

\begin{figure*}
	\centering
	\includegraphics[width=\hsize]{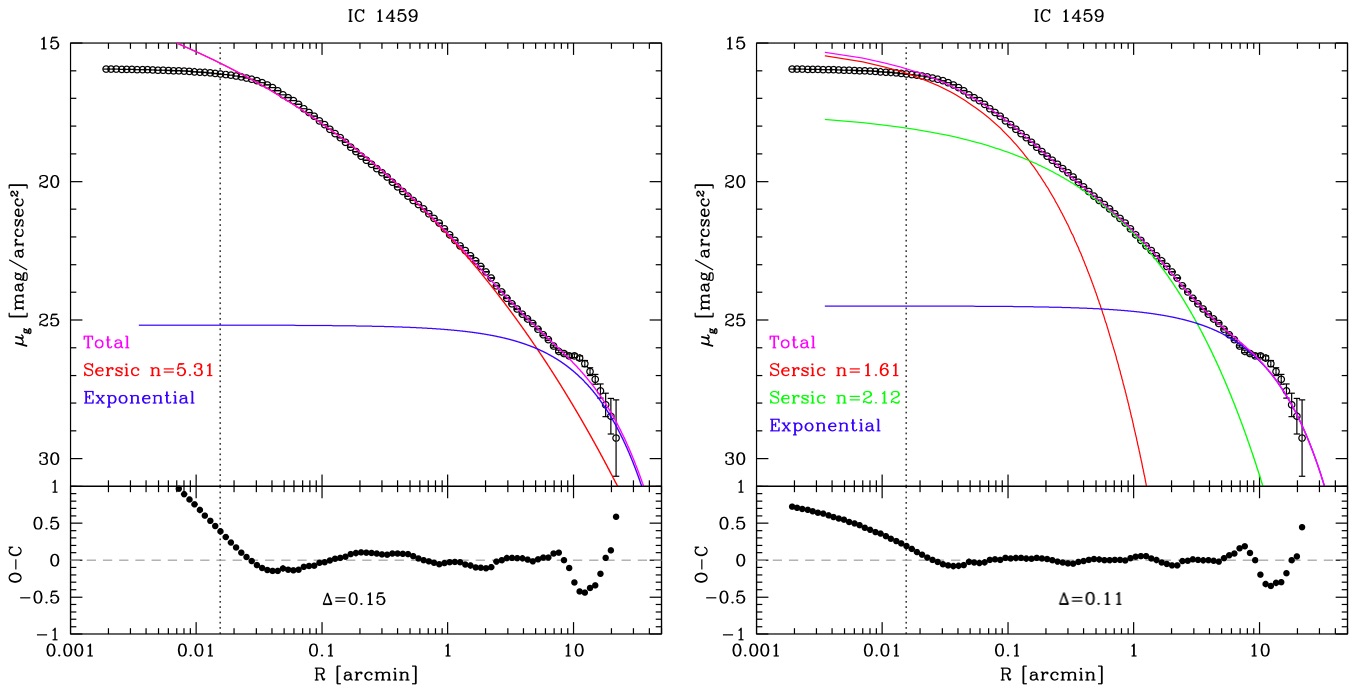}
	\includegraphics[width=\hsize]{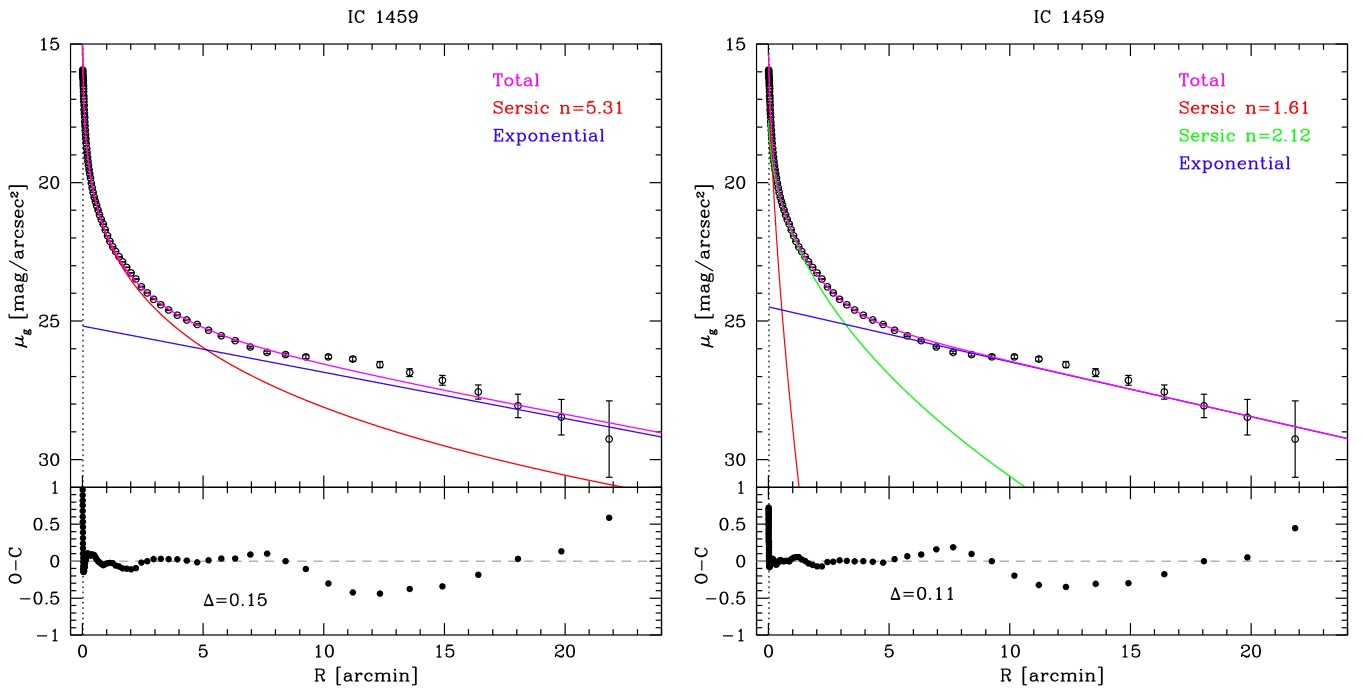}
	\caption{{\it g} band surface-brightness radial profile of IC~1459 (black open circles) on a logarithmic (top panels) and linear scale (bottom panels) fitted with a two-component model (left panels) and with a three-component model (right panels). The red solid line is the first Sérsic component, the green solid line is the second Sérsic component, the blue solid line is the exponential component, and the magenta solid line marks the model of the total light distribution. The black vertical dotted line marks the core of the galaxy which was excluded in the fit ($R<0\farcs9$). The $O-C$ panel (black filled circles) represents the residual between the azimuthally-averaged surface-brightness radial profile and the multi-component model. $\Delta$ is the rms scatter minimised by the Levenberg-Marquardt algorithm \citep{Seigar2007}.} 
	\label{fig:1459fit1d}
\end{figure*}

For this reason we have fitted the azimuthally-averaged surface-brightness radial profiles of IC~1459 combining a Sérsic law and an exponential function. 
From the total magnitude of the Sérsic and exponential component, $m_{\rm tot,1}$ and $m_{\rm tot,2}$ respectively, taking into account the stellar mass-to-light ratio derived from the average $g-r$ colours (see Tab.~\ref{tab:colmass}), 
{  we derived the relative contribution of the outer envelope with respect to the total stellar mass of the galaxy (it is $f_h$ in Table~\ref{tab:fit1d1459}) which is about the 37\%.} Results are listed in Table~\ref{tab:fit1d1459} and shown in Figure~\ref{fig:1459fit1d}.

Numerical simulations by \citet{Cooper2013,Cooper2015,Pillepich2018} 
suggest that the surface brightness profile of BCGs/BGGs could be described by the superimposition of three components. In these simulations, the first Sérsic law represents the central {\it in-situ} component; the second Sérsic law and the third exponential function represent the accreted component, the relaxed and un-relaxed respectively. 
As discussed in detail by \citet{Spavone2017a,Spavone2017b}, in order to compare observations and theoretical predictions, we {{  used numerical simulations as ``prior'' guide for the decomposition, and}} fitted the light distribution of IC~1459 also with a three-component model, where the inner S\'ersic component should "mimic" the in-situ component predicted by the above numerical simulations. 
The fitted parameters are listed in Table~\ref{tab:fit1d1459}. 
Figure~\ref{fig:1459fit1d} shows the results of the three-component fits. 
{{  Looking at the rms scatter $\Delta$ of the fit, we can clearly see that by adding the third component we achieve an improvement of at least 4\%. Since the expected value of $\Delta$ scales as $\sqrt{(m-k)/m}$ (see \citet{Seigar2007}), where $m$ is the number of measured points ($\sim$ 70 in our case) and $k$ is the number of free parameters, we would need 11 free parameters to obtain an improvement of 4\%. This means that the improvement we obtain in our fit is not only due to the introduction of additional free parameters, as already shown by \citet{Seigar2007} and \citet{Spavone2017b}.} }

From the three-component model, we have estimated the transition radius 
$R_{\rm tr}=3\farcm1 \simeq 25.7$~kpc, which is the distance from the galaxy centre at which the unrelaxed component starts to dominate the light distribution of the BGG \citep{Cooper2013}. 
From the two accreted stellar components, the relaxed and un-relaxed ones, described by the second Sérsic law 
and exponential function, we estimate the total accreted mass fraction of $87\%$ in IC~1459, which corresponds to 
$8.77 \times 10^{11}$~$\mathcal{M}_{\odot}$ (see Table~\ref{tab:bcg}).
This is the relative contribution of the accreted components with respect to the total stellar mass of the galaxy, 
$f_{h,T}$, to be compared with the predictions for this quantity by the numerical simulations cited above. The comparison is given in Fig.~\ref{fig:haloratio} and commented in Sec.~\ref{subsec:bcg}. {  The error bars on the datapoint have been derived by means of error propagation on $M_{\star}$ and $M_{acc}$, which are 0.02 and 0.04, respectively.}
Table~\ref{tab:bcg} reports the values of transition radius, mass-to-light ratio, total stellar mass, 
and accreted mass fraction of IC~1459.

%----------------------------------Table 4--------------------------------------------------
\begin{table*}
    \centering
    \scriptsize
    %\footnotesize
	\caption{Best fitting structural parameters for the multi-component fit of the surface-brightness radial profile of IC~1459 in the $g$ band.}
	\label{tab:fit1d1459}
	\begin{tabular}{lcccccccccccc}
		\hline\hline
		Models& 
		\multicolumn{2}{c}{$R_{\rm e,1}$}& 
		$n_1$& 
		$\mu_{\rm e,1}$& 
%		$m_{\rm tot,1}$& 
		\multicolumn{2}{c}{$R_{\rm e,2}$}& 
		$n_2$& 
		$\mu_{\rm e,2}$& 
%		$m_{\rm tot,2}$& 
		\multicolumn{2}{c}{$r_{h}$}& 
		$\mu_0$& 
%		$m_{\rm tot,3}$&
		$f_{h,T}$\\
		&
		\multicolumn{1}{c}{(arcsec)}&
		\multicolumn{1}{c}{(kpc)}&
		& 
		(mag~arcsec$^{-2}$)& 
%		(mag)& 
		\multicolumn{1}{c}{(arcsec)}&
		\multicolumn{1}{c}{(kpc)}&
		&
		(mag~arcsec$^{-2}$)& 
%		(mag)& 
		\multicolumn{1}{c}{(arcsec)}&
		\multicolumn{1}{c}{(kpc)}&
		(mag~arcsec$^{-2}$)& 
%		(mag)&
		\\
		(1)&
		\multicolumn{2}{c}{(2)}&
		(3)&
		(4)&
%		(5)&
		\multicolumn{2}{c}{(5)}&
		(6)&
		(7)&
%		(9)&
		\multicolumn{2}{c}{(8)}&
		(9)&
%		(12)&
		(10)\\
		\hline
		$2$ & $53.79\pm0.06$ & $7.15$ & $5.3\pm0.7$ & $21.7\pm0.5$ & $\cdots$ &$\cdots$&$\cdots$&$\cdots$ & $390\pm1$ & $52.61$ & $25.18\pm0.01$ & $37\%$\\
		$3$ &$5.54 \pm 0.12$ & $0.67$ & $1.61 \pm 0.06$ & $18.19 \pm 0.04$ & $54 \pm 1$ & $7.28$ & $2.12 \pm 0.12$ & $21.69 \pm 0.05$ & $329 \pm 4$ & $44.38$ & $24.49 \pm 0.11$ & $87\%$\\
%		{\it r} &$6.84 \pm 0.22$&$0.81$&$2.65 \pm 0.21$&$17.72 \pm 0.01$&$10.16$&$65.88 \pm 0.72$&$8.77$&$2.05 \pm 0.16$&$21.47 \pm 0.20$&$8.99$&$317.30 \pm 15.70$&$42.76$&$24.24 \pm 0.89$&$9.74$&$82\%$\\
%		{\it i} &$\pm$&$ $&$\pm$&$\pm$&$ $&$\pm$&$ $&$\pm$&$\pm$&$ $&$\pm$&$ $&$\pm$&$ $&$\%$\\
		\hline\hline
	\end{tabular}
	\tablefoot{Best fit parameters of the PSF-deconvolved light profiles decomposition: {\it Col.~1} - number of functions of the multi-component fit applied. {\it Col.~2}, {\it Col.~3}, and {\it Col.~4} - effective radius, Sérsic index, and effective surface brightness of the first Sérsic component. {\it Col.~5}, {\it Col.~6}, and {\it Col.~7} - effective radius, Sérsic index, and effective surface brightness of the second Sérsic component. {\it Col.~8}, and 
	{\it Col.~9} - scale radius, and central surface brightness of the exponential component. {\it Col.~10} - accreted mass fraction.}
\end{table*}
%------------------------end Table 4--------------------------------------------------

%---------------------------------------------- Figure 4------------------------------------
%---------------------------------------------- end Figure 4------------------------------------

\subsection{Intragroup Light}\label{subsec:1459igl}

{  As stated before, from a detailed inspection of the deep VST mosaic we do not detect any intragroup low surface 
brightness features in the form of stellar tails or streams between and/or around the group members, 
except for those in the outskirts of IC~1459 (see Sec.\ref{sec:morphology} and Fig.~\ref{fig:1459SB}). 
Therefore, we consider this area as the bulk of IGL and focused on it to estimate the total flux in this component.
To this aim, we have modelled the light distribution in the {\it g} band of the dominant member IC~1459 with the IRAF task \textsc{bmodel}. 
This task creates a two-dimensional noiseless photometric model of the galaxy from the result of the 
isophotal analysis generated by \textsc{ellipse}, by taking into account the variations in the 
ellipticity and position angle. We have derived the residual image by subtracting the two-dimensional model from the parent image.
All the foreground and background sources in the residual image were masked. In particular,
for the  three brightest stars HD~216666 in the area ($\alpha_{J2000}=22^{\rm h}55^{\rm m}14^{\rm s}.947$ and $\delta_{J2000}=-36^{\circ}23\arcmin19\farcs18$), HD~216781 ($\alpha_{J2000}=22^{\rm h}56^{\rm m}12^{\rm s}.13$ and $\delta_{J2000}=-36^{\circ}40\arcmin49\farcs3$), and HD~216972 ($\alpha_{J2000}=22^{\rm h}57^{\rm m}50^{\rm s}.33$ and $\delta_{J2000}=-36^{\circ}32\arcmin45\farcs8$)\footnote{The stars coordinates are from \textsc{HyperLeda}}, masks extend out to regions where scattered light is detected (see Fig.~\ref{fig:1459SB}).
The IGL region is defined with the IRAF task \textsc{polymark} and it is about 23 arcmin$^2$ around 
IC~1459. From the fit of the light profiles, we know that the stellar envelope starts to dominate at
$R\geq R_{\rm tr}=3\farcm1=25.77$ kpc (see Sec.~\ref{subsec:1459ld}). 
Therefore, for the IGL estimate we account for the flux from $R\geq R_{\rm tr}$, and the inner regions of the galaxy (at smaller radii) are masked.

From the defined regions, we have derived the integrated extinction-corrected magnitude of the IGL.
The total luminosity of the IGL in {\it g} band is $5.25 \times 10^9$~L$_{\odot}$, 
that accounts for the $2\%$ of the total light of the group and $3\%$ of the light of IC~1459 (see Table~\ref{tab:iglHI}).
The error estimate on the flux in the selected area takes into account all sources that contributed to the residual fluctuations in the sky background, as given in Sec.~\ref{sec:oss}.
}

\section{A deep view of the IC~1459 Group}\label{sec:1459groupana}

Deep OmegaCAM@VST data allow us to map the {  azymuthally-averaged} surface brightness of 
IC~1459 down to $\mu_g= 29\pm1$~mag~arcsec$^{-2}$ and $\mu_r= 29\pm1$~mag~arcsec$^{-2}$ at 
$R=21\farcm83\simeq181.1$~kpc ($\sim7.4 R_{\rm e}$ in {\it g} and $\sim9.8 R_{\rm e}$ in {\it r} band) (see right 
panel of Figure~\ref{fig:1459profiles}).
Images in the $i$ band are shallower (see Tab.~\ref{tab:obs-log} and Sec.~\ref{sec:oss}), 
which cause the observed drop of the surface brightness profile in this band (see Fig.~\ref{fig:1459profiles}).
{  IC~1459 is the biggest and the reddest galaxy of the group, with $R_{{\rm e},g} \simeq 23.9$~kpc and $g-r = 0.84$~mag (Table~\ref{tab:mtotre}). 
By fitting the light distribution we found that the outer stellar envelope starts to dominates the light for $R\geq 5$~arcmin (see Sec.~\ref{subsec:1459ld} and Tab.~\ref{tab:fit1d1459}).
The other group members are LTGs, except for IC~5269 that is classified as an S0
(see Fig.~\ref{fig:mosaic}). Main properties of LTGs (i.e. total luminosity, effective radii and average colours) are given 
in Table~\ref{tab:mtotre} and Table~\ref{tab:colmass}.

The enlarged  region  of  the  VST  mosaic around IC~1459 in the g-band surface brightness levels (see 
Fig.~\ref{fig:1459SB}) highlights the structure of the IC~1459 outer envelope (i.e. at $R\geq 5$~arcmin). 
There are shells in the North-East and South-West regions and prominent sharp edges on the west side.
We detected an elongated tidal tail on the south, and thick arc-like tail protruding on the West side of the small
group member IC~5264.
The isophotal analysis shows that the outer envelope has a twist in the position angle profile of about 40 degrees 
and a scatter in the ellipticity profile being roundish (see left panel of Figure~\ref{fig:1459profiles}). }

{  From the averaged colour profile (Figure~\ref{fig:1459col}, left panel), inside 1~arcmin ($\sim 8$~kpc)
IC~1459 has redder colours, where $g-r$ varies in the range $0.5 - 1.5$~mag. This is the region 
where \citet[][ and references therein]{Forbes1995} detected an arcsec-scale dust lane crossing the galaxy nucleus. 
At larger radii, the colour profiles decline toward bluer colours, with $g-r\leq0.5$~mag.
The two-dimensional colour map (Figure~\ref{fig:1459col}, right panel) shows an asymmetric distribution of the colour: 
in the north, there is a arc-like structure, extending from East to West, characterised by very red colours ($g-r\geq0.6$~mag). This structure corresponds to the region where shells are detected in the envelope.
In the region of the tidal tail (in the south) the envelope has bluer colours ($g-r\simeq0.6$~mag). 
The small peculiar LTG ESO~406-27 in the group, which is located in this region, is the bluest galaxy of the group ($g-r\sim0.34$, Table~\ref{tab:colmass}). The existence of the blue tidal tails close to ESO~406-27 might 
confirm the previous hypothesis that there is a possible ongoing interaction of this galaxy with IC~1459 \citep{Serra2015,Saponara2018}.}

Figure~\ref{fig:mosaic} shows the VST {\it g} band mosaic of the group with the HI map from \citet{Oosterloo2018}. There is a 
lot of HI associated to LTGs, showing a different degree of asymmetry and off-centre distribution respect to the stellar disk. 
In the region of the BGG there are signatures of HI stripping from gas-rich galaxies \citep{Saponara2018}, 
like the north-east to south-west low surface-brightness HI tail \citep{Oosterloo2018}. 
The possible HI donors of these debris are IC~5264 and ESO~406-27 \citep{Saponara2018,Oosterloo2018}.
{  In the NE of IC~1459, the HI seems not associated with galaxies: the deep VST images shows that this is the region where 
shells in the envelope are detected. 
In the southern part of the IC~1459 envelope, the HI distribution is also elongated EW, overlapping the region where
the faint tidal tail is found (see Fig.~\ref{fig:1459SB}).
Furthermore, more in the south, we found a faint diffuse 
light over-density north to ESO~406-27 (see Fig.~\ref{fig:eso40627sp}), 
which would be consistent with  the interaction scenario proposed by \citet{Serra2015}, \citet{Saponara2018} and \citet{Oosterloo2018} which involves IC~5264 and ESO~406-27.

In conclusion, the faint features detected in the IC~1459's envelope  provide evidence of the on-going accretion 
process on the BGG. Furthermore,  the interaction and tidal effects could be responsible of HI stripped debris and of 
the faint optical counterparts found in two group members \citep{Kilborn2009,Serra2015,Saponara2018}.}

\section{IC~1459 versus other Loose Groups of Galaxies}\label{sec:lgg}

In this section we compare the main observed properties of the IC~1459 group with those available for the other two 
groups of galaxies centred on NGC~5018 \citep{Spavone2018} and NGC~1533 \citep{Cattapan2019}. 
{  The comparison is motivated by consistency in the data set and in the total mass of the systems. 
All groups are targets of the VEGAS sample. Therefore, all data were obtained with the same telescope and 
with the same observing strategy, so they have comparable depth and accuracy. Furthermore, for 
the data analysis (i.e. isophote fit, fitting of the surface brightness profiles, colour estimate) we adopted 
the same tools and procedures described in Sec.~\ref{sec:sp}.
In addition, the three groups have the virial mass of the same order of magnitude. 
NGC~5018 and NGC~1533 groups have comparable virial radius and mass, with $R_{vir}=0.4$~Mpc and
$M_{vir}\sim7 \times 10^{12}$~M$_{\odot}$ for NGC~5018, and $R_{vir}=0.4$~Mpc and
$M_{vir}\sim5 \times 10^{12}$~M$_{\odot}$ for NGC~1533 \citep{Gourgoulhon1992,Firth2006}, 
whereas IC~1459 is a massive system, with $R_{vir}=0.21$~Mpc and $M_{vir}\sim 3.7 10^{13}$~M$_{\odot}$ \citep[see Sec.~\ref{subsec:1459group} and][]{Brough2006}.}

In the following sections, we compare the characteristics of the BGGs, the stellar halo properties, 
the amount of IGL, and the HI mass and distribution. This helps to trace the different evolutionary stages and how this could reflect in the observed properties.

\begin{figure}
	\centering
	\includegraphics[width=\hsize]{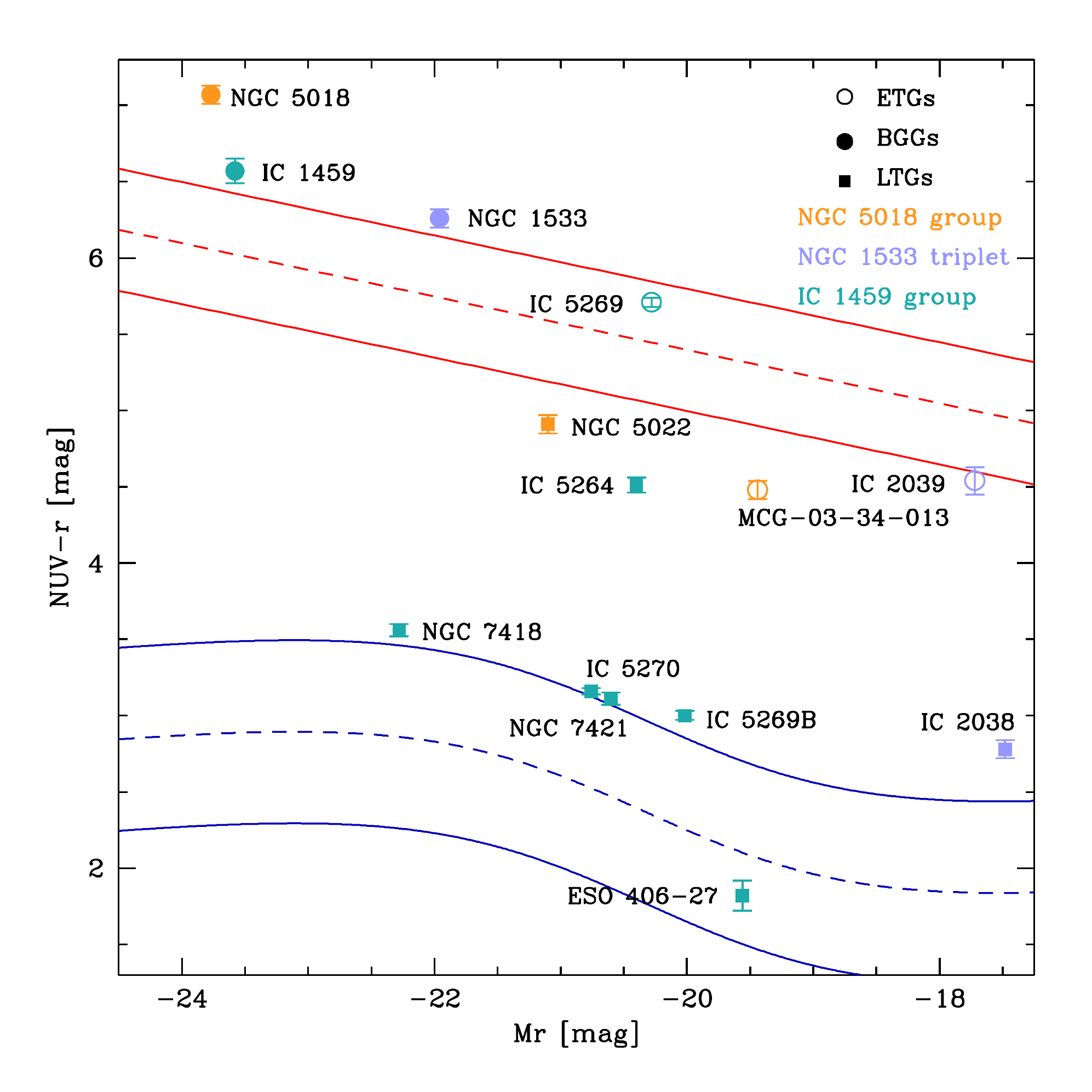}
	\caption{M$_r$ vs. ($NUV-r$) colour–magnitude diagram of the three loose groups: the NGC~5018 group (orange points), the NGC~1533 triplet (mauve points), and the IC~1459 group (teal points). The \citet{Wyder2007} fits (dashed lines), plus the error estimations (solid lines), to the Red Sequence (red lines) and Blue Cloud (blue lines) are shown. Open circles mark early-type galaxies (ETGs), filled circles mark the brightest group galaxies (BGGs), and filled squares mark late-type galaxies (LTGs). The NUV data are from \texttt{NED}, while the {\it r} band data for the NGC~5018 group are from \citet{Spavone2018}, for the NGC~1533 triplet are from \citet{Cattapan2019}, and for the IC~1459 group are presented in this work.}
	\label{fig:cmd}
\end{figure}

An overall and preliminary picture of the evolutionary stage of the groups is carried out here using the color-magnitude diagram (CMD). Here it is possible to trace the galaxy transformation from active and star-forming (Blue Cloud, BC) to passively (Red Sequence, RS) evolving system via the Green Valley \citep[GV, see e.g.][and references therein]{Mazzei2014}. 
Figure~\ref{fig:cmd} shows the CMD of the three analysed low density environments.
The BGG of the NGC~5018 group, NGC~5018, seems to be a typical red and dead ETG, and the other two galaxies (NGC~5022 and MCG-03-34-013) are in the GV approaching the RS.
{  The very red NUV-r colours for NGC~5018 could be reasonably due to the large amount of dust in the galaxy centre \citep{Spavone2018}, which affects the NUV flux. }
In the NGC~1533 triplet the BGG, NGC~1533 is in the RS, while the other ETG (IC~2039) is very close to the RS, and the LTG (IC~2038) has just left the BC. 
The CMD of the entire backbone of the Dorado group, including the NGC~1533 triplet,
has been shown by \citet{Cattapan2019}. The Dorado backbone has an extended and rich RS.
IC 1459 and IC 5269 are the two ETGs of the IC~1459 group and they lie in the RS, as expected. IC 5264 is the LTG that lies in projection of the stellar halo of IC~1459 and it is in the GV. All the other LTGs, except for ESO~406-27 which is in the BC, are leaving the BC to migrate in the GV. 

This means that the IC 1459 group is a typical young galaxy group with a significant BC and a depopulated RS. And on the other hand the NGC~5018 group represents what happens for more evolved galaxy systems which have a well defined RS and a poor or empty BC \citep[][ and references therein]{Cattapan2019}.

%---------------------------------------------- Figure ?------------------------------------
%---------------------------------------------- end Figure ?------------------------------------

\subsection{The accreted mass fraction of BGGs}\label{subsec:bcg}

\begin{figure}
	\centering
	\includegraphics[width=\hsize]{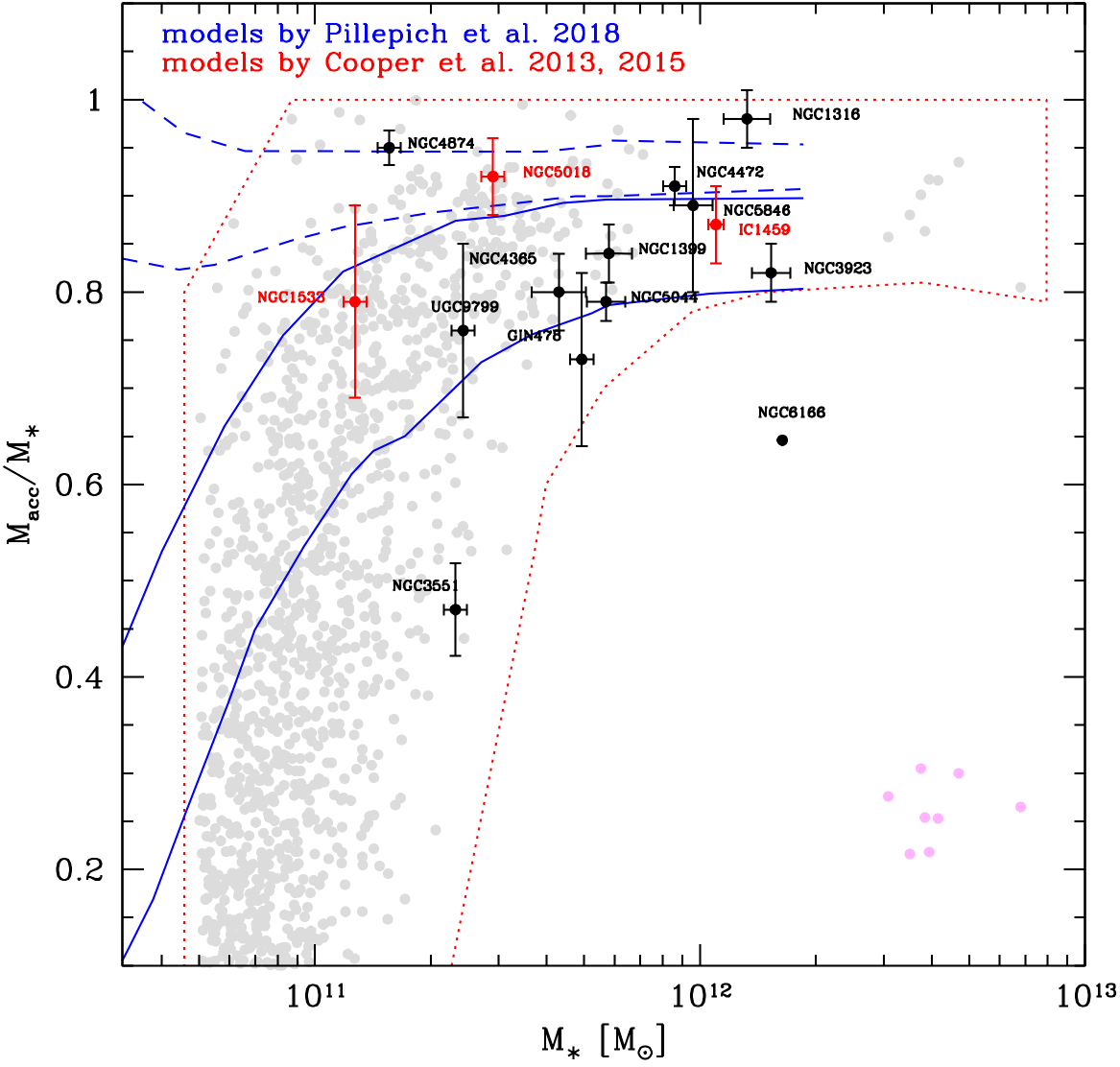}
	\caption{Accreted mass fraction as a function of total stellar mass for ETGs. The measurements for NGC~5018, NGC~1533, and IC~1459 (from the three-components fit) are given as red circles. Black circles correspond to other BGGs/BCGs from the literature \citep{Seigar2007,Bender2015,Iodice2016,Iodice2017a,Spavone2017b,Spavone2018}. The region within the red dashed lines and the grey filled circles correspond to the predictions of cosmological galaxy formation simulations by \citet{Cooper2013,Cooper2015}. The regions brackets by the blue continuous and dashed lines indicate the accreted mass fraction measured in the Illustris simulations by \citet[][; see their Figure~$12$]{Pillepich2018} within $30$~kpc and outside $100$~kpc, respectively. Magenta filled circles show the mass fraction associated with the streams from table~1 in \citet{Cooper2015}.}
	\label{fig:haloratio}
\end{figure}

The three BGGs we analyse are NGC~5018, NGC~1533 and IC~1459. 
IC~1459 is the brightest ($L^* _{\rm tot} = 1.77 \times 10^{11}$~$L_{\odot}$), more massive ($\mathcal{M}^* _{\rm tot} = 10.08 \times 10^{11}$~$\mathcal{M}_{\odot}$) and reddest galaxy ($g-r = 0.84$~mag). While NGC~1533 is the less luminous ($L^* _{\rm tot} = 2.99 \times 10^{10}$~$L_{\odot}$) and less massive galaxy ($\mathcal{M}^* _{\rm tot} = 2.9 \times 10^{11}$~$\mathcal{M}_{\odot}$). NGC~5018 is the bluest galaxy ($g-r = 0.7$~mag). The properties of the BGGs are listed in Table~\ref{tab:bcg}.
In Figure~\ref{fig:haloratio} we show the accreted mass fraction as a function of the total stellar mass for the analysed 
BGGs/BCGs in VEGAS \citep{Iodice2016,Iodice2017a,Spavone2017b,Spavone2018,Cattapan2019} and for the other BGGs/BCGs in the 
literature \citep{Seigar2007,Bender2015}. We compare these results with the theoretical predictions of cosmological galaxy 
formation  by \citet{Cooper2013,Cooper2015} and with the Illustris simulations by \citet[][their Figure~$12$]{Pillepich2018}. 
We find that the accreted mass fraction {  for the three BGGs (IC~1459, NGC~5018 and NGC~1533) 
is in the range of $78\% - 92\%$ (Table~\ref{tab:bcg}). These values are comparable with those obtained for other BCGs and BGGs
in VEGAS sample, as well as with literature data (see Fig.~\ref{fig:haloratio}). Moreover, the accreted mass fraction estimated
for IC~1459 and previously for NGC~5018 and NGC~1533 is consistent with the theoretical predictions, which suggest that 
stars accreted by BCGs/BGGs in the stellar mass range $10^{11}-10^{13}$ account for most of their total stellar mass \citep{Cooper2013,Pillepich2018}.

The above result also suggests that the accreted stellar mass fraction seems to be a function of the total stellar mass rather 
than of the environment, since the BGGs have comparable accreted mass  to the bright cluster members 
(Figure~\ref{fig:haloratio}). } This lack of correlation between accreted mass fraction and environment is a very recent studied 
topic and it is developed for BCGs by \citet{DeMaio2018}.

%---------------------------------------------- Table 5------------------------------------
\begin{table*}
    \centering 
    \small
	\caption{Main properties of the BGGs sample.}
	\label{tab:bcg}
	\begin{tabular}{lcccccccccc}
		\hline\hline
		BGG&
		$g-r$&
		$(M/L)_g$&
		$\mathcal{M}^* _{\rm tot}$&
		$f_{\rm h,T}$&
		$\mathcal{M}^* _{\rm tot}$~$_{\rm acc}$&
		$R_{\rm e}$&
		$R_{\rm tr}/R_{\rm e}$&
		$\mu_{\rm tr}$&
		$(g-r)_{R<R_{\rm tr}}$&
		$(g-r)_{R>R_{\rm tr}}$\\
		&
		(mag)&
		&
		($\times 10^{11} \mathcal{M}_{\odot}$)&
		&
		($\times 10^{11} \mathcal{M}_{\odot}$)&
		(kpc)&
		&
		(mag~arcsec$^{-2}$)&
		(mag)&
		(mag)\\
		(1)&(2)&(3)&(4)&(5)&(6)&(7)&(8)&(9)&(10)&(11)\\
		\hline
		NGC~5018 &$0.70\pm0.20$ & $1.97$ & $2.9$ & $92\%$ & $2.7$&$6.37$&$5.37$&$26.6$&$0.78$&$0.94$\\
		NGC~1533 & $0.77\pm0.05$ & $3.92$ & $1.17$ & $78\%$ &  $0.91$& $6.35$& $2.80$& $27.2$& $0.77\pm0.04$& $0.49\pm0.29$\\
		IC~1459& $0.84\pm0.31$ & $5.70$ & $10.08$ & $87 \%$&  $8.77$& $24.47$& $1.05$& $25.1$& $0.89\pm0.29$& $0.44\pm0.11$\\
		\hline\hline
	\end{tabular}
	\tablefoot{{\it Col.~1} - BGGs name. {\it Col.~2} - averaged extinction-corrected $g-r$ colour value. {\it Col.~3} - mass-to-light ratio in {\it g} band. {\it Col.~4} - total stellar mass. {\it Col.~5} and {\it Col.~6} - total accreted stellar mass fraction from the multi-component fit in {\it g} band and value in solar masses. {\it Col.~7} - Effective radius in {\it g} band. {\it Col.~8} and {\it Col.~9} - transition radius normalised to the effective radius and corresponding surface brightness. {\it Col.~10} and {\it Col.~11} - averaged extinction-corrected $g-r$ colour value inside and outside the transition radius.}
\end{table*}
%-------------------------- end Table 5----------------------------------------------------------
%---------------------------------------------- Figure 6------------------------------------
%---------------------------------------------- end Figure 6------------------------------------

\subsection{The stellar envelope of the BGGs}\label{subsec:sh}

\begin{figure}
	\centering
	\includegraphics[width=\hsize]{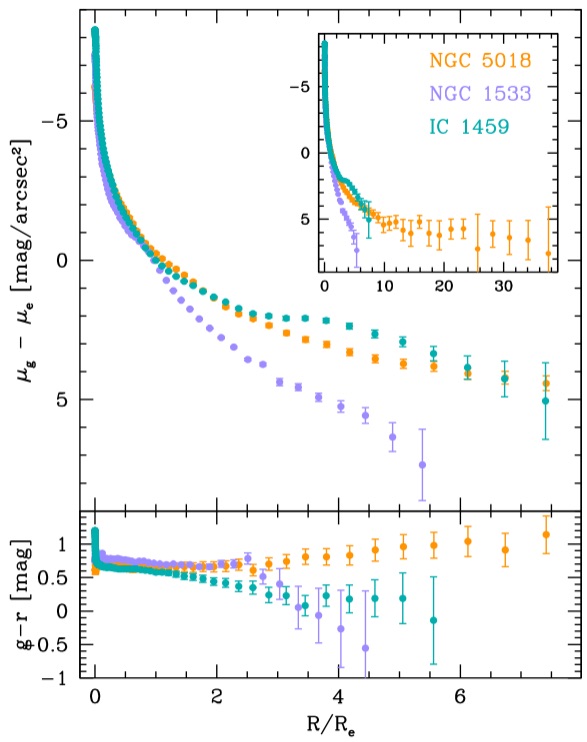}
	\caption{Azimuthally-averaged surface-brightness (top panel), and colour normalised (bottom panel) radial profiles of NGC~5018 (orange), NGC~1533 (mauve) and IC~1459 (teal) in {\it g} band scaled to their effective surface brightness, $\mu_{\rm e}$, as a function of the semi-major axis normalised to the effective radius, $R/R_{\rm e}$.}
	\label{fig:sh}
\end{figure}

{  By fitting the light distribution, we were able to set the scales of the different components in the galaxy structure.
In particular, we estimated the transition radius where the outer stellar and faint envelope starts to dominate the light (see Sec.~\ref{subsec:1459ld}).
The stellar envelopes of the three BGGs studied in this work occur at surface brightness levels $25.1<\mu_g<27.2$~mag~arcsec$^{-2}$
(see Table~\ref{tab:bcg}),
which is comparable with the surface brightness of the stellar envelopes observed in other BCGs 
\citep[][$24.0<\mu_g<27.8$~mag~arcsec$^{-2}$]{Spavone2017a} and even consistent with the theoretical predictions.
In the three BGGs, the transition in the surface brightness profiles also corresponds to a transition in the ellipticity, 
position angle and colour profiles. In particular, the outer isophotes are more elongated or rounder than the inner 
ones and are twisted \citep[see Fig.~\ref{fig:1459profiles} and][]{Spavone2017a,Spavone2018,Cattapan2019}. 

In Figure~\ref{fig:sh} we have compared the azimuthally-averaged surface brightness (in the $g$ band) and colour profiles
for the three BGGs, scaled to their effective magnitude, as a function of $R/R_{\rm e}$. 
This reveals that NGC~5018 has the most extended envelope, out to 30 R$_e$, which is also quite red 
$(g-r)_{R>R_{\rm tr}} = 0.94$~mag.
IC~1459 and NGC~1533 have smaller ($\sim 10$~R$_e$) and bluer envelopes, with $(g-r)_{R>R_{\rm tr}} = 0.44$~mag and 
$(g-r)_{R>R_{\rm tr}} = 0.49$~mag, respectively.
At radii larger than $R_{\rm tr}$ (i.e. $\geq 2$~R$_e$), the surface brightness profiles show that the contribution to the total
light from the stellar envelope in NGC~5018 and IC~1459 is larger than that in NGC~1533. This is consistent with an higher accreted mass fraction estimated in the former two galaxies (see Tab.~\ref{tab:bcg}).

According to the theoretical predictions on the mass assembly on the BCGs and BGGs, the morphology of the outskirts, the shape of the light profiles and colour distribution reflect the different accretion processes and progenitors  \citep{Cooper2010,DSouza2014,Monachesi2019,Mancillas2019}. The gradual accretion of small-mass satellites produces streams, 
the intermediate and major merging generates shells and tidal tails. In the outskirts of three BGGs, the deep VEGAS images
have shown such kind of features, suggesting that the mass assembly is still ongoing.
The most prominent and luminous tidal tails are observed in the NGC~5018 group, in the intra-group space and protruding from
 the BGG \citep{Spavone2018}. In IC~1459 (this work) and in NGC~1533 we observe very faint tidal tail, 
which are probably tracing the interaction with the smaller group members close to the BGG, and several shells in the outskirts \citep{Cattapan2019}.  

According to \citet{Mancillas2019},  the evolution of the number of tidal tails does not change with the age of the BCG/BGG, 
 the expected number is from 1 to 3 at maximum and have a short survival time of $\sim 2$~Gyr.
Stellar streams show a peak in number (reaching 8 to 10) around 10~Gyr of the galaxy age, 
decreasing in number for older galaxies, 
since they tend to be dissolved in the halo.  
Shells strongly depend on the inclination, therefore the prediction on the detectable 
number changes from 4 to 8 for a galaxy with an age of 10~Gyr. 
Both streams and shells are longer-lived than tidal tails, surviving until 4~Gyr.
Given that, and taking into account that the last burst of star formation for NGC~5018 is about 4~Gyr \citep{Spavone2018}, 
whereas in IC~1459 and NGC~1533 it is older (around 10~Gyr) the expected number of tidal tails is consistent with the
observations in the same range of surface brightness levels. 
The number of shells and streams observed in IC~1459 and NGC~1533 is also consistent with simulations in the range of age
estimated for these two galaxies. The absence, or few faint, streams and shells in NGC~5018 is also expected 
from simulations for galaxies of comparable age.

Summarising, the difference or similarity in the global properties (light profiles and substructures) 
of the stellar envelope in three BGGs might constrain the phase and/or the mechanism in the mass assembly. 
In particular,  NGC~5018 might have experienced  strong tidal forces in the latest 2 Gyr during the interaction with the 
companion bright galaxy, which lead to the prominent tidal tail. Differently, since the IC~1459 and NGC~1533 groups are
populated by less luminous galaxies close to the BGG, their stellar halo is assembling by minor and intermediate merging,
which shaped the observed shells and streams. }

\subsection{Intragroup Light}\label{subsec:igl}

{  In Figure~\ref{fig:iglHI} (top panel) we compare the fraction of IGL with
respect to the total light in the group, as function of the virial mass, 
available for several groups of galaxies, including the
estimates we derived for three systems discussed in this work.
Although the sample is composed of eight groups in total, the large scatter suggests that there is
no clear trend of IGL with the virial mass.
For massive cluster of galaxies ($M_{vir} \geq 10^{13}$), this result would be consistent with 
theoretical predictions from \citet[][]{Contini2014} and \citet[][]{Rudick2011} 
where the relations between ICL and virial mass is quite flat. 
On the other hand, the different IGL values found in the three groups is consistent with 
previous observations indicating that the higher the ETGs/LTGs ratio, the 
greater the IGL component is \citep{DaRocha2008}.
As pointed out by \citep{DaRocha2008}, the low IGL fraction is expected for 
groups that are LTGs dominated and have HI still in the discs of the galaxies.
The NGC~5018 group and the NGC~1533 triplet are composed of two ETGs and one LTGs \citep{Spavone2018,Cattapan2019}. 
While the IC~1459 group counts only two ETGs, IC~1459 and IC~5269, i.e. 
the ETGs-to-LTGs ratio is quite low, ETGs/LTGs$=0.29$ (Table~\ref{tab:iglHI}).
The IGL component of the NGC~5018 group is the highest ($\sim 41\%$) between the three group studied here. 
A smaller fraction of IGL is derived for the other two systems, it is 
$\sim 8\%$ for NGC~1533 triplet and $\sim2\%$ for IC~1459 group (Table~\ref{tab:iglHI}).
Compared to the other two groups, IC~1459 has a large amount of HI associated to the LTGs group members.
All others groups of galaxies included in Fig.~\ref{fig:iglHI} are Hickson 
Compact Group (HCG) of galaxies, which are usually characterised by an high ETGs/LTGs 
ratio \citep{Hickson1992}. 
The IGL fraction obtained for NGC~5018  \citep{Spavone2018} is similar to 
the IGL estimated for HCG~90 \citep{White2003}, which is a strongly interacting system 
with $L_{\rm IGL} / L_{\rm Group} \sim 38 \%$ of comparable virial mass.
The HCG88 is the only system of the HCG sample 
that shows no IGL component, down to the surface brightness detection limit \citep{DaRocha2008}.  

The above comparison suggests that, even considering the different detection 
limit of the observations and the slightly different approach to estimate the 
intra-group diffuse light, the IGL estimate for the three groups analysed in 
this work is consistent with previous values for groups of galaxies of comparable
virial mass. Values also agree with fractions of ICL quoted in the literature, which ranges 
from 10 to 40\% going from groups to clusters \citep[e.g.][]{Feldmeier2004,Zibetti2005,McGee2010,Toledo2011}.

The fraction of diffuse light in groups and cluster predicted from numerical simulations also spans the same 
range of values. Predictions from \citet{Sommer-Larsen2006} are $\sim 12\% - 45\%$, while more recently \citet{Contini2014} found $\sim 10\%-40\%$.
The amount of ICL depends on the formation mechanisms. 
About 5 to 25\%  of  the  diffuse  light builds up by the infalling galaxies 
in the potential well of the BCG/BGG during the mass assembly history.
Theoretical works predict that the bulk of the ICL is produced by the most  massive  satellite  
galaxies, M$\sim 10^{10 - 11}$~M$_\odot$ \citep{Purcell2007,Contini2014,Martel2012}. 
The contribution to the diffuse light from lower-mass galaxies (M $\leq 10^9$~M$_\odot$) is very little, 
even if they are more numerous. 
Therefore, the low amount of IGL detected in the IC~1459 group might be 
connected to the absence in the group of a comparable mass galaxy that is interacting with the BGG (see Tab.~\ref{tab:colmass}). 

}

%---------------------------------------------- Table 6------------------------------------
\begin{table*}
	\centering
	\caption{Properties of galaxy group sample.}
	\label{tab:iglHI}
		\begin{tabular}{lccccccccc}
			\hline\hline
		Group&N& ETGs/LTGs&$m_{{\rm IGL},g}$&$L_{{\rm IGL},g}$&$( L_{\rm IGL}/L_{\rm BGG}) _g$&$( L_{\rm IGL}/L_{\rm Group})_g$&$\mathcal{M}_{\rm HI}$&$D$\\
		&&& (mag)& $( \times 10^{9} L_{\odot} )$ &&&$( \times 10^{9} M_{\odot} )$ & (Mpc)\\
		(1)&(2)&(3)&(4)&(5)&(6)&(7)&(8)&(9)\\
			\hline
			NGC~5018 &$3$&$1.6$&$11.39$&$70.60$&$47\%$&$41\%$&$2.4$&$31.4$\\
			NGC~1533 &$3$&$1.6$&$14.17$&$3.46$&$12\%$&$8\%$&$7.7$&$21.0$\\
			IC~1459 &$9$&$0.3$&$13.10$&$7.17$&$4\%$&$2\%$&$22.2$&$28.7$\\
			\hline\hline
		\end{tabular}
	\tablefoot{{\it Col.~1} - group name. {\it Col.~2}  and {\it Col.~3} - number of bright galaxies in the group ($M_g < -17$) and fraction of ETGs. {\it Col.~4} and {\it Col.~5} - integrated magnitude and luminosity in {\it g} band of the IGL. {\it Col.~6} and {\it Col.~7} - fraction of intragroup light divided by the BGG luminosity and total luminosity of the group. {\it Col.~8} - total HI mass associated to the group. {\it Col.~9} - distance of the group used to estimate the total HI mass; for NGC~5018 by \citet{Kim1988}, for NGC~1533 by \citet{Kilborn2005} and for IC~1459 by \citet{Kilborn2009}.}
\end{table*}
%-------------------------- end Table 6----------------------------------------------------------

\subsection{HI distribution versus diffuse light}\label{subsec:HI}

\begin{figure}
	\centering
	\includegraphics[width=\hsize]{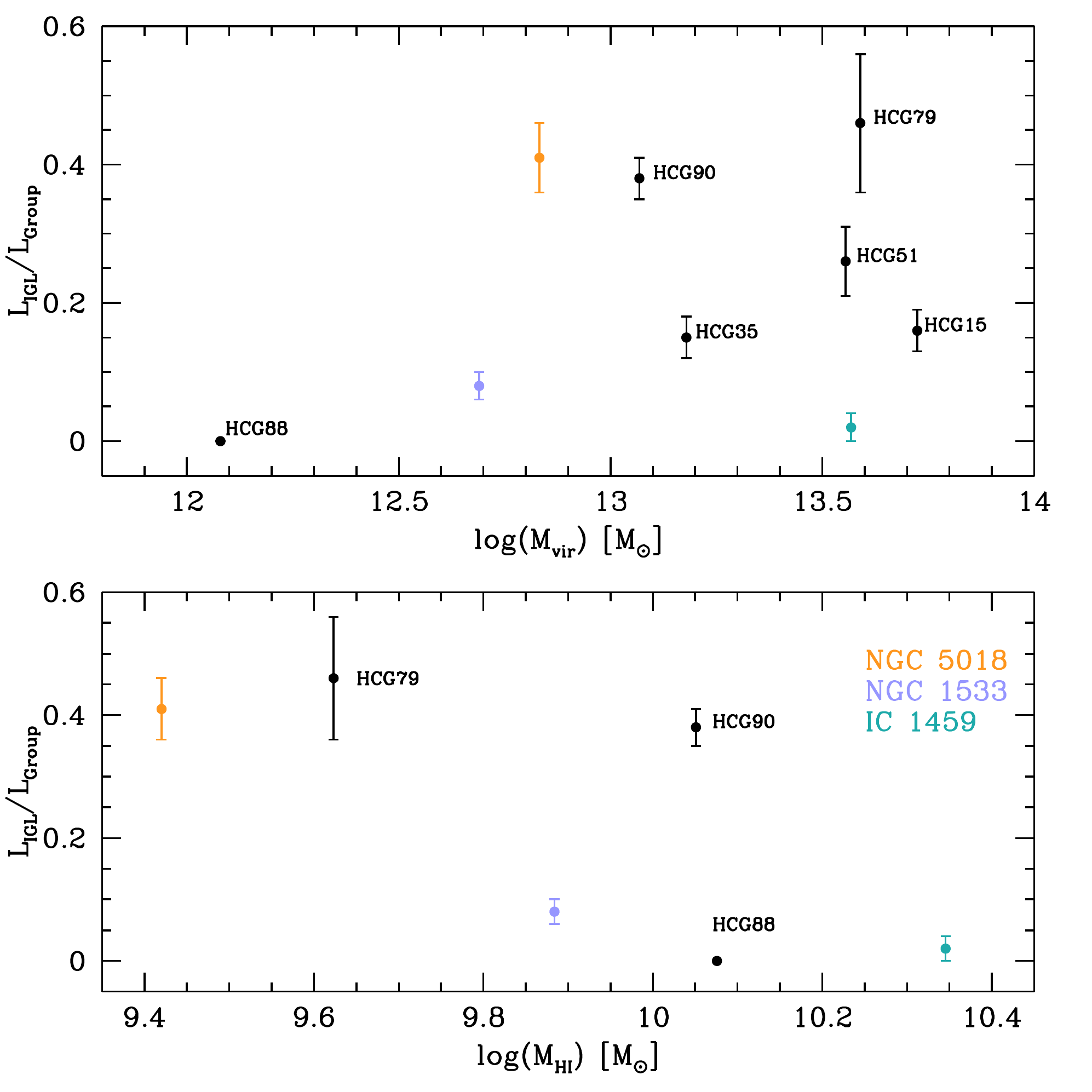}
	\caption{  Luminosity of the IGL component normalised to the total group luminosity as a function of the total HI mass of the group (lower panel) and of the virial mass of the group (top panel), for NGC~5018 group (orange), NGC~1533 triplet (mauve), IC~1459 group (teal). Values are compared with those for several Hickson Compact groups (HCGs). For HCG~79 and HCG~88 data are from \citet{Borthakur2010,DaRocha2005,Ribeiro1998,Nishiura2000}. For the remaining HCGs data are taken from \citet{DaRocha2008,Selim2008}}.
	\label{fig:iglHI}
\end{figure}

{  In this section, for the three groups studied in this work, we aim at comparing the HI distribution around the group members and in the intra-group space with the location of the LSB features detected.
The HI distribution in the NGC~5018 group and in NGC~1533 triplet results from
the ongoing tidal interactions between group members, which are stripping 
the cold gas from the LTGs \citep{Spavone2018,Cattapan2019}.
The debris of tidal stripping are located in the outskirts of the BGGs, i.e. NGC~5018 and NGC~1533, in the form of arc-like structures or extended tails.
From the deep VEGAS data, we found the optical counterpart for some of
the intra-group HI features, at very faint levels $\mu_g \simeq 28-30$~mag~arcsec$^{-2}$.

The IC~1459 group shows a different HI distribution. 
The HI is mainly associated with the group members and it follows the distribution of galaxies along a thick filament in the North-South direction \citep{Kilborn2009,Saponara2018}. 
In addition, Figure~\ref{fig:mosaic} shows that there
are few HI over-densities that seem not to be associated at any optical feature.
The most prominent are close to IC~1459, on the SE and on the NE side.
As pointed out in Sec.~\ref{subsec:1459igl}, the SE over-density might be associated
with the stellar faint tail detected on the south of IC~1459, and 
linked to the ongoing interaction involving IC~5264 and ESO~406-27 \citep{Saponara2018,Oosterloo2018}.

A clear difference between the IC~1459 group and the other two groups, NGC~5018
and NGC~1533, also arises by comparing the IGL component with the total HI mass 
of the group (i.e. galaxies and intragroup HI).
The fraction of IGL light decreases as the HI mass increases, see 
Fig.~\ref{fig:iglHI} (lower panel) and Table~\ref{tab:iglHI}.
Figure~\ref{fig:iglHI} suggests that this trend, i.e. lower amount of 
intra-cluster diffuse light is found in groups
with higher content of HI gas, seems also confirmed for other groups of galaxies
(with available HI and IGL measurements).

To conclude, by correlating the HI distribution and content with the LSB features 
and IGL amount, the IC~1459 group seems to be in a different evolutionary  
phase with respect to the NGC~5018 and NGC~1533 groups.
The low amount of IGL, the large HI content and its regular distribution might 
indicate that the IC~1459 group is still assembling. 
This would be in agreement with \citet{Saponara2018}, who suggested that, 
given the small HI velocity gradient in the south-southeast direction of IC~1459,
in combination with the high number of gas-rich members and low velocity dispersion, 
the group could be in a first stage of evolution.}

\section{Summary and Conclusions}\label{sec:conc}

In this work we have presented and analysed new deep imaging data from the VEGAS survey for the galaxy group IC~1459. 
The VST mosaics in {\it g}, {\it r} and {\it i} bands cover the whole group extension over an area of $1\times 2$~square degrees.
{  We have compared the properties of this group with those of other two low density systems from VEGAS, with similar virial masses: the NGC~5018 group and the NGC~1533 triplet. 
The NGC~5018 and NGC~1533 groups have similar environments: they have the same number of 
large galaxies and of ETGs, their HI mass has the same order of magnitude, 
and their galaxies are closer in projection to each other. 
On the contrary, the IC~1459 group is a richer environment with 9 bright galaxies of which 
7 are LTGs, it has a total HI mass of $\sim 10^{10}$~$\mathcal{M}_{\odot}$, and galaxies are located in projection along a filamentary structure 2~degrees long. 

This work aim at studying low-density 
environments, still unexplored at the faintest surface brightness levels 
where the stellar envelope in the galaxy outskirts and the intra-group light start to 
dominate. Taking advantage of the deep imaging from VEGAS, 
the main goal is to address the mass assembly history of the group and their members.
This is done by studying {\it i)} the structure of the BGG outskirts, where the "tracers" 
of the ongoing accretion (i.e. tidal tails, stellar streams and shells) can be detected; 
{\it ii)}  the azimuthally-averaged surface-brightness radial profile, in order to estimate
the accreted stellar mass component; {\it iii)} the amount of IGL and its distribution. 
The main properties listed above were correlated with the HI gas mass and distribution.

The main results of this work are the following.
	\begin{enumerate}
	    \item In the outskirt of IC~1459, which is the BGG of the group, we detected several LSB features in the surface brightness level 
	    $25 \leq \mu_g \leq 27$~mag~arcsec$^{-2}$. They are shells, with red colours, in the NE side and two faint and bluer tails in south (see Fig.~\ref{fig:1459SB}). 
	    All seem to be signs of interaction and accretion in the galaxy stellar halo. Same features are detected in the BGG of another group from the VEGAS sample, NGC~1533, from \citet[][]{Cattapan2019}. Differently from both of them, in the NGC~5018 group a prominent and very extended tidal tail was found, and the BGG's outskirt is characterised by several stellar streams \citep{Spavone2018}.
	    \item Like in NGC~5018 and NGC~1533, the azimuthally-averaged surface brightness profiles show an extended exponential envelope down to $\mu_g \sim 29$~mag/arcsec$^2$ and out about 9R$_e$ (see Fig.~\ref{fig:1459profiles}). By fitting the light distribution, we estimate that the accreted stellar mass in this galaxy is 87\%, which is similar to those derived for  NGC~5018 and NGC~1533 (92\% and 78\%, respectively) and to those of galaxies with comparable total stellar mass in other groups or clusters of galaxies (see Fig.~\ref{fig:haloratio}).
	    \item The IC~1459 group has a very low ($\sim2\%$) fraction of IGL compared to NGC~5018 and NGC~1533, consistently with its small ETGs/LTGs ratio and the high HI amount (see Fig.~\ref{fig:iglHI}).
	    
	\end{enumerate}

The above results suggest that the three groups are in a different phase of mass assembly.
The filamentary distribution of galaxies in IC~1459, where the HI gas is still associated 
to the 7 LTGs, and very low amount of IGL are an indication that there were few and minor
interactions between the group members and the BGG, 
which generated the shells and tidal tails detected in the outskirts of IC~1459 (also probably related to the HI over-density observed in the same regions). According to simulations \citep{Mancillas2019}, this would have happened in the latest 4 Gyr, 
which is the survival time of such a features.  
On the contrary, in both NGC~5018 and NGC~1533 groups, many more interactions have 
happened in their formation history that induced gas-depletion, by ram pressure stripping, and produced the higher IGL amount and the 
disturbed HI distribution which clearly traces the ongoing interaction between some group 
members \citep{Cattapan2019,Spavone2018}.
In fact, the HI distribution is related to the environment and to the interaction history 
of the galaxies. Typically, in interacting or merging galaxies in groups, the HI is located along tails, streams and bridges, indicating that these structures are formed by tidal 
stripping \citep{Bekki2005,Kilborn2009}.

The above scenario is consistent with a different evolutionary stage of the NGC~5018 and 
NGC~1533 groups with respect to IC~1459: both of them have an higher ETGs/LTGs ratio with 
respect to that in IC~1459. A larger amount of IGL is expected for more evolved systems 
with higher the ETGs/LTGs ratio \citep{DaRocha2008}.
The color-magnitude diagram also confirm that NGC~5018 and NGC~1533 groups are more evolved systems 
than IC~1549 group, since almost all group members are approaching the red sequence, 
while in IC~1459 group most of the members are still in the region of active and star 
forming galaxies (see Fig.~\ref{fig:cmd}).

As concluding remark, this work shows that the structure of the outer envelope of 
the BCGs (i.e. the signatures of past mergers and tidal interactions), 
the IGL component and the HI amount and distribution may be used as indicators of the different 
evolutionary stage and mass assembly in galaxy groups.
We plan to perform the analysis presented in this work on a larger group sample by VEGAS project in the next two years.
}

%\begin{acknowledgements}
      {\it Acknowledgements}$-$This work is based on visitor mode observations taken at the ESO La Silla Paranal Observatory within the VST Guaranteed Time Observations, Programme IDs 097.B-0806(B), 098.B-0208(A) and 0100.B-0168(A). {  The authors wish to thank the anonymous referee for his or her comments and suggestions that allowed us to greatly improve the paper. EI acknowledge financial support from the ESO inside the visitor program 2019/2020.}
      AC, MS and EI acknowledge financial support from the VST project (P.I. P. Schipani). The VST project is a joint venture between ESO and the National Institute for Astrophysics (INAF) in Naples, Italy. RR acknowledges funding from the INAF PRIN-SKA 2017 program 1.05.01.88.04. EMC acknowledges financial support from Padua University through grants DOR1715817/17, DOR1885254/18, DOR1935272/19, and BIRD164402/16. GD acknowledges support from CONICYT project Basal AFB-170002.
%\end{acknowledgements}

%\clearpage

\bibliographystyle{aa.bst}
  \bibliography{1459}
  
%\clearpage

\appendix

\section{Surface Photometry of IC~1459 group}\label{apx:A}

\subsection{IC~5270}
This late-type barred spiral galaxy is the northernmost galaxy of the group, lying at a projected distance of $37\farcm3$ from IC~1459. From the VST images (Figure~\ref{fig:ic5270sp}) there is not a clear optical counterpart for the north-northeast HI asymmetric distribution. The edge-on galaxy inclination does not allow us to investigate the stellar disk asymmetries and bar component.

\subsection{IC~5269}
This is a barred lenticular galaxy with an average colour ($g-r=0.77\pm0.25$~mag) consistent with the predicted colour of ETGs \citep{LaBarbera2012}. It is the forth most massive ($3.32 \times 10^{10}$~$\mathcal{M}_{\odot}$) and smallest bright galaxy ($R_{{\rm e},g}=2.70$~kpc) galaxy of the group. From the surface brightness profiles there is a clear evidence for the bulge and stellar disk components. Its colour profile is almost flat outside of the seeing-dominated region.
It does not have any associated HI (Figure~\ref{fig:ic5269sp}). 

\subsection{IC~5269B}
This is the second closest galaxy in projection to IC~1459 ($14\farcm5$). It is the second less bright, $M_g=-19.60$~mag, and less massive, $5.0 \times 10^{9}$~$\mathcal{M}_{\odot}$, bright galaxy of the group.
This galaxy is very inclined with a bluer outer regions (Figure~\ref{fig:ic5269Bsp}). There is an off-centre of the HI component respect to the stellar disk, that seems to be more elongated in the opposite direction. 
This could be due to some galaxy-galaxy or galaxy-environment interaction, providing a tidal stripping component and this gas might build-up the debris structure around IC~1459.

\subsection{IC~5264}
This is a peculiar, very small ($R_{{\rm e},g}=4.05$~kpc), edge-on, early-type spiral. The average colour ($g-r=0.79\pm0.06$~mag), is too red for a LTG, but this galaxy is seen in projection in the IC~1459 stellar halo. Hence this could explain that they have a similar $g-r$ average colour. From the VST images (Figure~\ref{fig:ic5264sp}), we were able to detect the warped structure of the outer stellar envelope and a dust lane in the southeaster side of the stellar disk in agreement with the HI off-centre distribution, which seems to be tangent to the outer envelope of IC~1459 in the eastern direction. Its HI component is faint and it is spread in a wide velocity range \citep{Serra2015}.
It has been suggested that IC~5264 is the principal donor of HI gas around IC~1459 due to a tidal interaction \citep{Saponara2018}.

\subsection{ESO~406-27}
This is the bluest ($g-r = 0.34\pm0.20$~mag) bright galaxy of the group, and has a total stellar mass ($3.2 \times 10^{9}$~$\mathcal{M}_{\odot}$) one order of magnitude smaller than the other galaxies except for IC~5269B.
It is located on the South-West side of IC~1459, at a project distance of $19.5\arcmin$. It is one of the HI brightest sources with a HI mass of $( 4.6 \pm 0.9) \times 10^9$~$\mathcal{M}_{\odot}$ \citep{Serra2015}.
From Figure~\ref{fig:eso40627sp}, ESO~406-27 might be interacting both with IC~1459 and NGC~7418. It has a two spiral-arms pattern clearly in the direction of these galaxies, which corresponds to the HI asymmetry. However we have not enough information yet for making an hypothesis regarding the evolution/interaction scenario between IC~1459, ESO~406-27, and NGC~7418.
\textsc{GALEX} data confirms that there are two spiral, UV-bright arms extending from northeast to southwest \citep{Thilker2007}.

\subsection{NGC~7418}
It is the second brightest ($M_g=-21.98$~mag), massive ($10.87 \times 10^{10}$~$\mathcal{M}_{\odot}$), and biggest ($R_{{\rm e},g}=21.72$~kpc), galaxy of the group and it is at a projected distance of $34\farcm8$ from IC~1459 in the South region of the group.
This late-type spiral has a very asymmetric stellar disk in the North-West direction. It has a weak optical bar, seen in the surface brightness profile rather than in the optical colour composite image (Figure~\ref{fig:ngc7418sp}). It seems to be stronger in {\it H} band imaging \citep{Eskridge2002}. The spiral arms are well defined in the inner regions, becoming bluer and smoothed in the outer parts.
As explored by \citet{Serra2015} and \citet{Oosterloo2018}, the complex HI tidal features elongated from the North region of the group to NGC~7418 seems to be the result of the first tidal galaxy-galaxy and galaxy-environment interactions having occurred in the BGG neighbourhood.

\subsection{NGC~7421}
It is a late-type barred spiral with an inner resonance ring \citep{Buta1995}. It has the smallest relative velocity ($v_{\rm rel} = -9.59$~km~s$^{-1}$) with respect to IC~1459. 
It is the third smallest, $R_{{\rm},g}=5.26$~kpc, and less massive, $2.44 \times 10^{10}$~$\mathcal{M}_{\odot}$, galaxy.
The colour composite VST image (Figure~\ref{fig:ngc7421sp}, left panel) shows the small inner bar and the South-East light asymmetries due to the spiral arms and the outer envelope of the galaxy, that are consistent with the off-centre of the HI disk. 
%It is a HI-deficient galaxy \citep{Denes2014} with $\mathcal{M}_{\rm HI}=(1.1\pm0.3) \times 10^{9}$~$\mathcal{M}_{\odot}$ \citep{Serra2015}.
The recent study made by \citet{Serra2015} has shown that the HI asymmetry in the opposite direction of the star-forming optical disk might be the signature of ram pressure stripping. 
According to \citet{Denes2014}, this galaxy is a HI-deficient galaxy, $\mathcal{M}_{HI}=(1.1\pm0.3) \times 10^{9}$~$\mathcal{M}_{\odot}$ \citep{Serra2015}.
However according to \citet{Ryder1997} and \citet{Ryder2000}, the tenuous hot intragroup medium is not able to produce a significant ram pressure stripping and hence tidal interactions might produce a couple of tails. The conclusion is that the galaxy had a previous tidal interaction, maybe with NGC~7418, which affected the stellar disk and the HI gas, and a subsequent weak ram pressure stripping, as happened for IC~5273, which dislocated the HI gas.

\subsection{IC~5273}
This late-type barred spiral is the third brightest galaxy of the group ($M_g=-21.59$~mag). It is the most distant galaxy in projection from IC~1459 ($79\farcm2$), and it has the larger relative velocity to IC~1459 ($v_{\rm rel}=-509 \pm 17$~km~s$^{-1}$). The bar is clearly visible (Figure~\ref{fig:ic5273sp}) both in the colour composite image and in the surface-brightness radial profile ($6\arcsec \leq R \leq 40\arcsec$). 
The average colours are $g-r=0.54$~mag and $g-i=1.05$~mag, and the colour profiles have a bluer decline for $R>3\arcsec$.
It is also one of the brightest sources in ASKAP HI, $\mathcal{M}_{\rm HI}=(5.4\pm1.1) \times 10^{9}$~$\mathcal{M}_{\odot}$ \citep{Serra2015}. There is an off-centre between the HI gas and stellar disk, and the HI distribution has a south-east asymmetry opposite to the direction of IC~1459.
The galaxy could have suffered some interactions with the environment or from ram pressure stripping, as for NGC~7421 \citep{Serra2015}. 

\section{Surface photometry}\label{apx:B}

\begin{figure*}
	\centering
	\includegraphics[width=18cm]{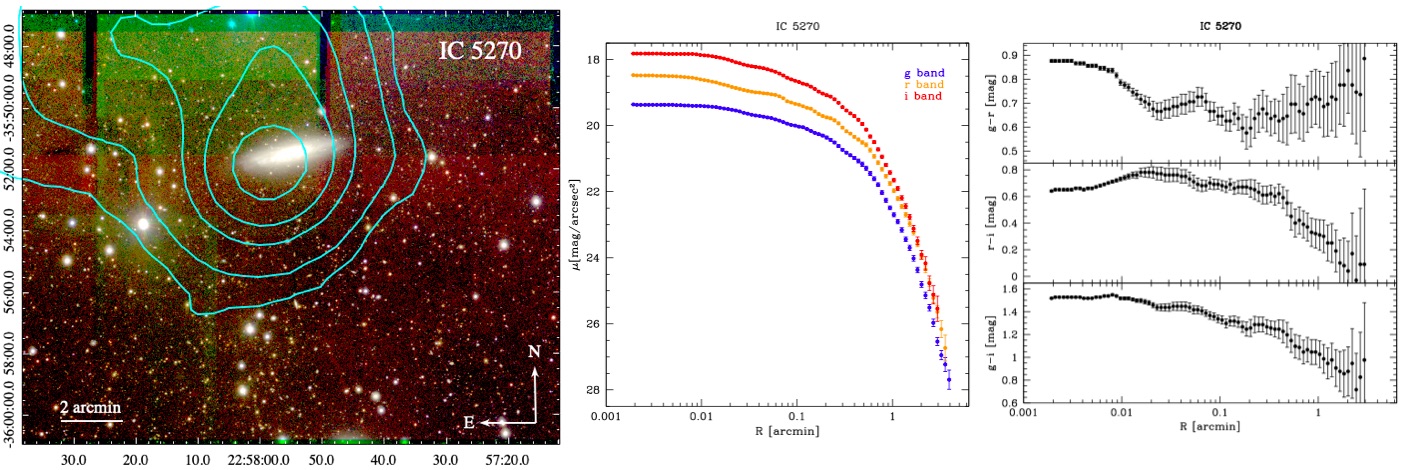}
	\caption{Left panel: colour composite image of IC~5270 with north at the top and east on the left.
    The right ascension and declination (J2000) are given in the horizontal and vertical axis of the field of view, respectively. The HI contours levels (white contours) are $1$, $2.5$, $7.5$, $12.5$, and $20 \times 10^{19}$~cm$^{-2}$. The white arrow indicates the direction of the BGG, IC~1459, respect with the galaxy. 
    Middle panel: {\it g}, {\it r}, and {\it i} band azimuthally-averaged surface-brightness radial profile.
    Right panel: azimuthally-averaged extinction-corrected $g-r$, $r-i$, and $g-i$ colour profile.}
	\label{fig:ic5270sp}
\end{figure*}

\begin{figure*}
	\centering
	\includegraphics[width=18cm]{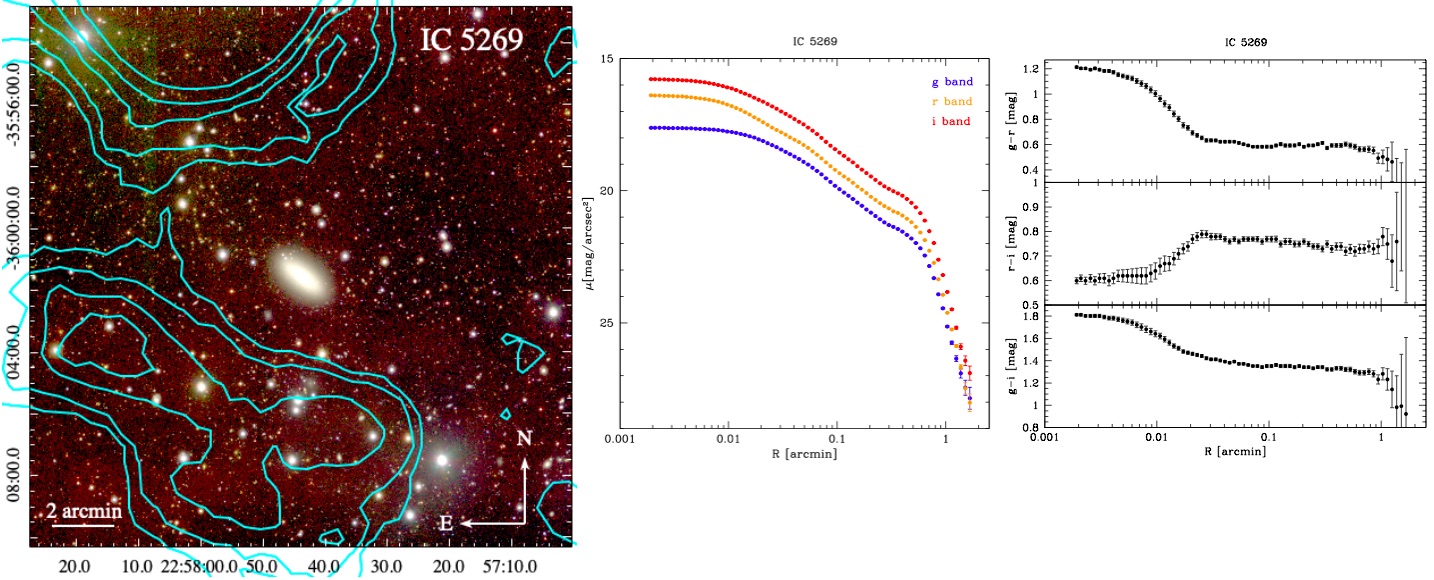}
	\caption{Same as Figure~\ref{fig:ic5270sp} but for IC~5269. The HI contour levels are $0.25$, $0.5$, $1$, $1.5$, and $2.5 \times 10^{19}$~cm$^{-2}$.}
	\label{fig:ic5269sp}
\end{figure*}

\begin{figure*}
	\centering
	\includegraphics[width=18cm]{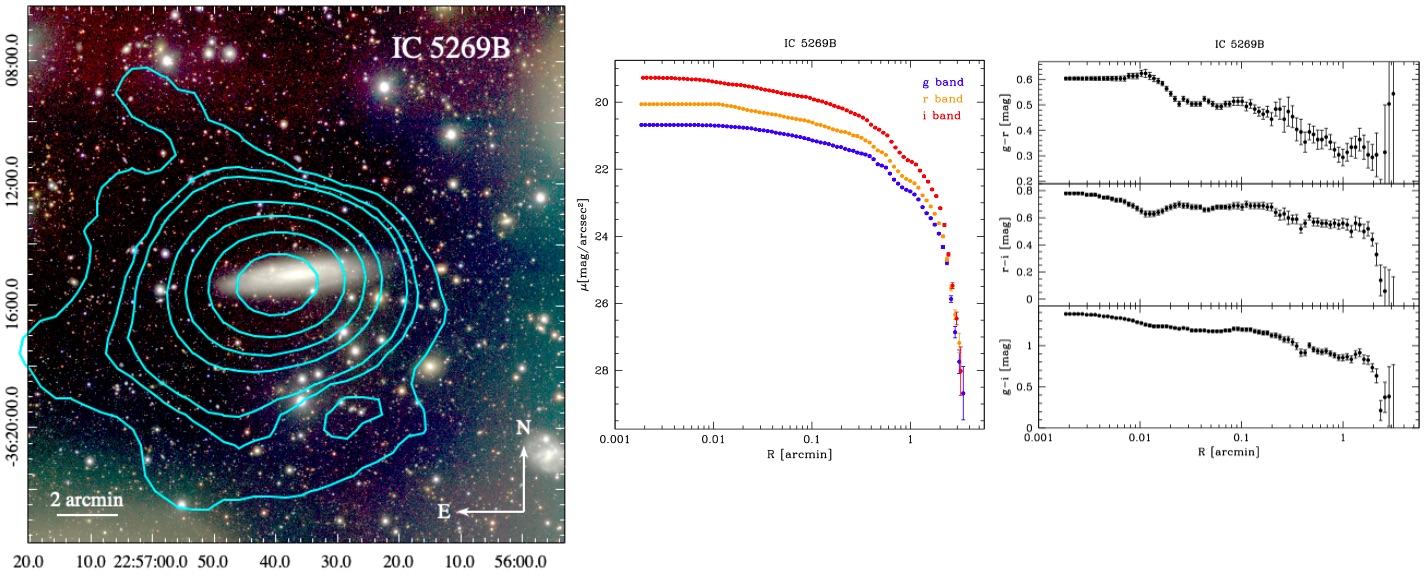}
	\caption{Same as Figure~\ref{fig:ic5270sp} but for IC~5269B. The HI contour levels are $0.25$, $1$, $2.5$, $7.5$, $12.5$, $20$, and $30 \times 10^{19}$~cm$^{-2}$.}
	\label{fig:ic5269Bsp}
\end{figure*}

\begin{figure*}
	\centering
	\includegraphics[width=18cm]{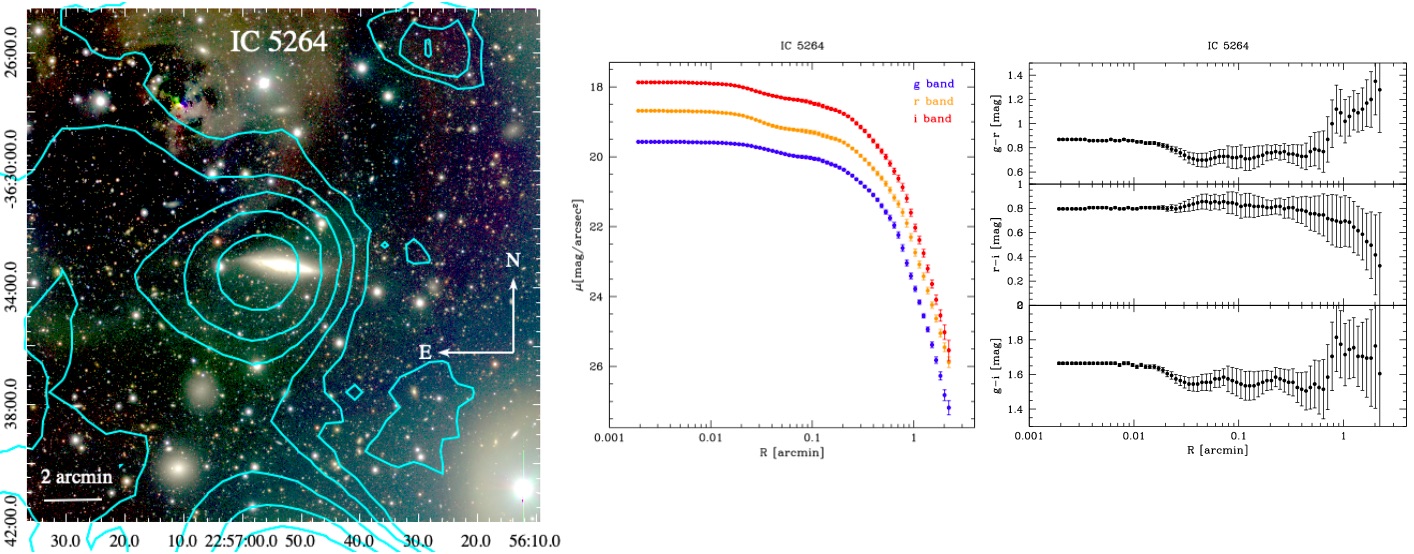}
	\caption{Same as Figure~\ref{fig:ic5270sp} but for IC~5264. The HI contour levels are $0.25$, $1$, $2.5$, $5$, and $10 \times 10^{19}$~cm$^{-2}$.}
	\label{fig:ic5264sp}
\end{figure*}

\begin{figure*}
	\centering
	\includegraphics[width=18cm]{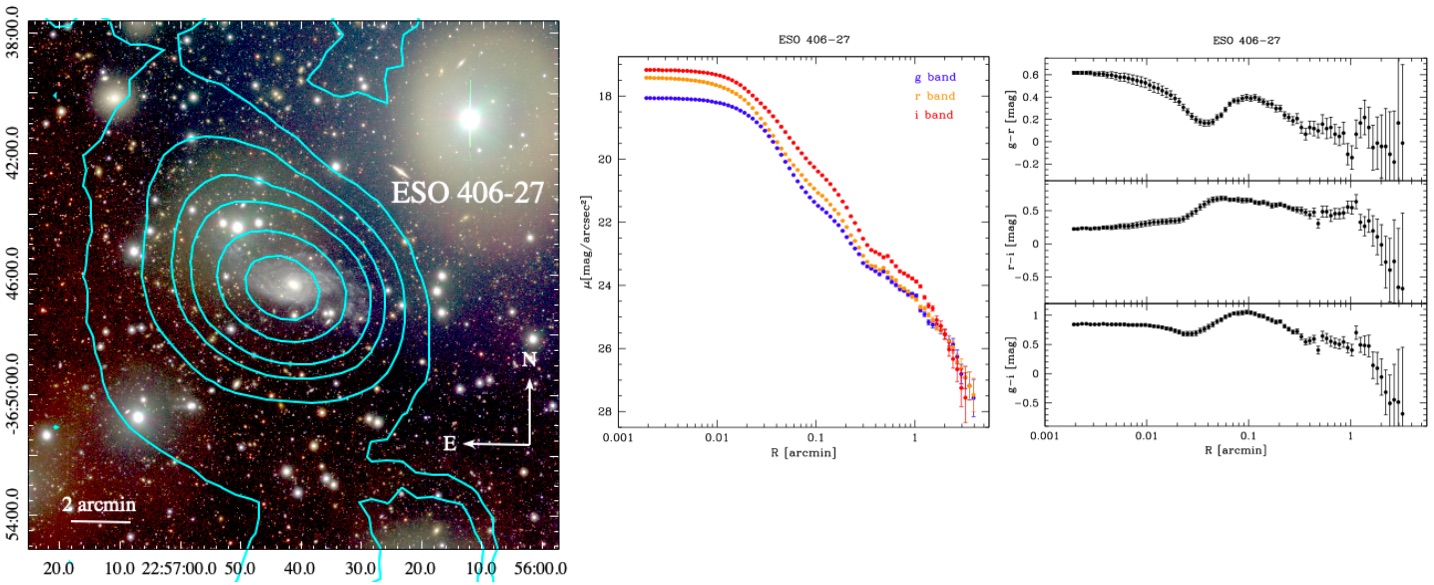}
	\caption{Same as Figure~\ref{fig:ic5270sp} but for ESO~406-27. The HI contour levels are $1.5$, $5$, $10$, $15$, $25$, and $35 \times 10^{19}$~cm$^{-2}$.}
	\label{fig:eso40627sp}
\end{figure*}

\begin{figure*}
	\centering
	\includegraphics[width=18cm]{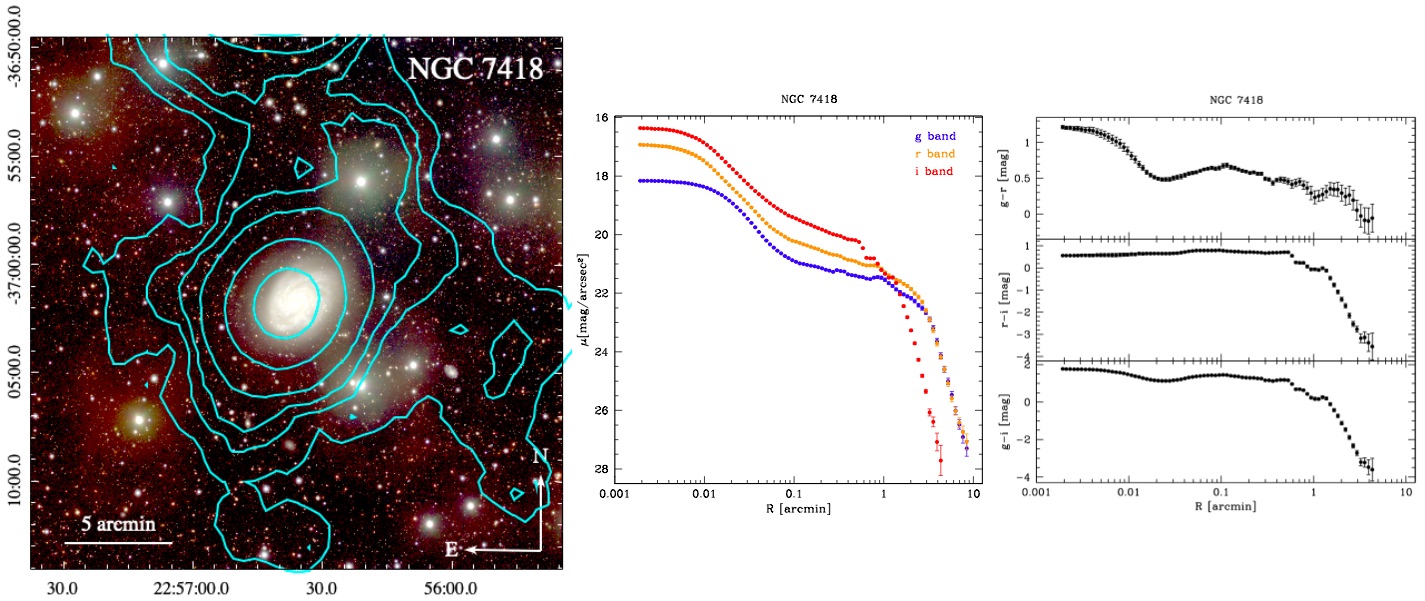}
	\caption{Same as Figure~\ref{fig:ic5270sp} but for NGC~7418. The HI contour levels are $0.5$, $1.5$, $2.5$, $5$, $15$, and $35 \times 10^{19}$~cm$^{-2}$.}
	\label{fig:ngc7418sp}
\end{figure*}

\begin{figure*}
	\centering
	\includegraphics[width=18cm]{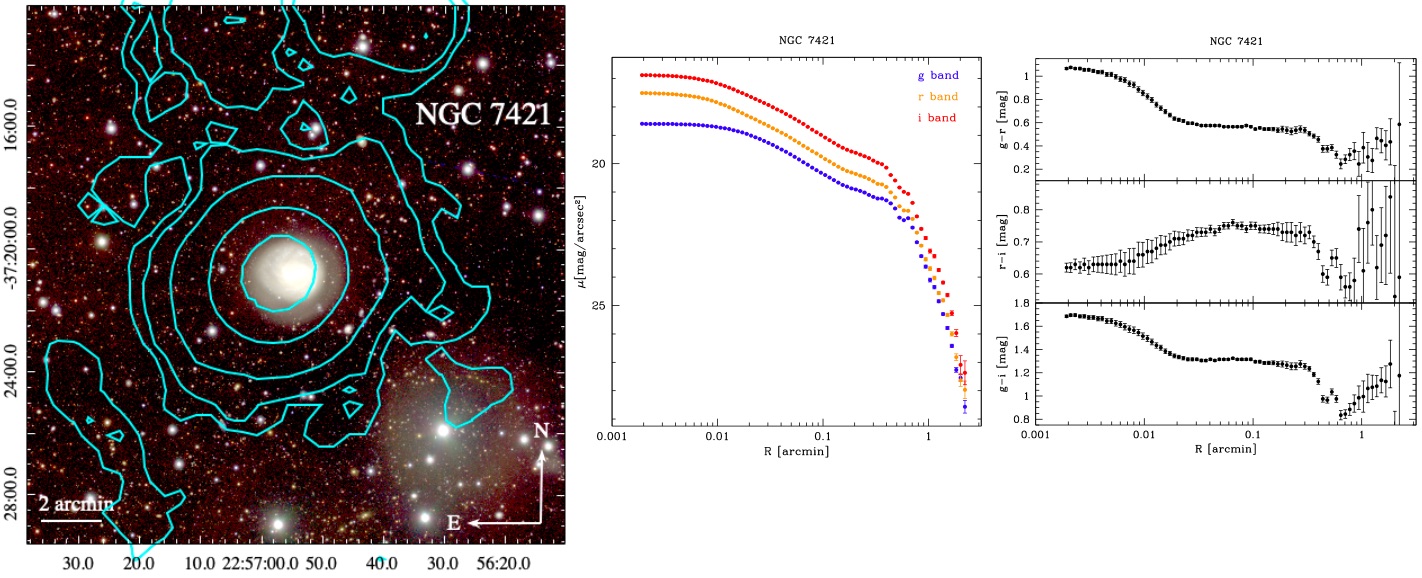}
	\caption{Same as Figure~\ref{fig:ic5270sp} but for NGC~7421. The HI contour levels are $0.25$, $0.5$, $1.5$, $5$, and $10 \times 10^{19}$~cm$^{-2}$.}
	\label{fig:ngc7421sp}
\end{figure*}

\begin{figure*}
	\centering
	\includegraphics[width=18cm]{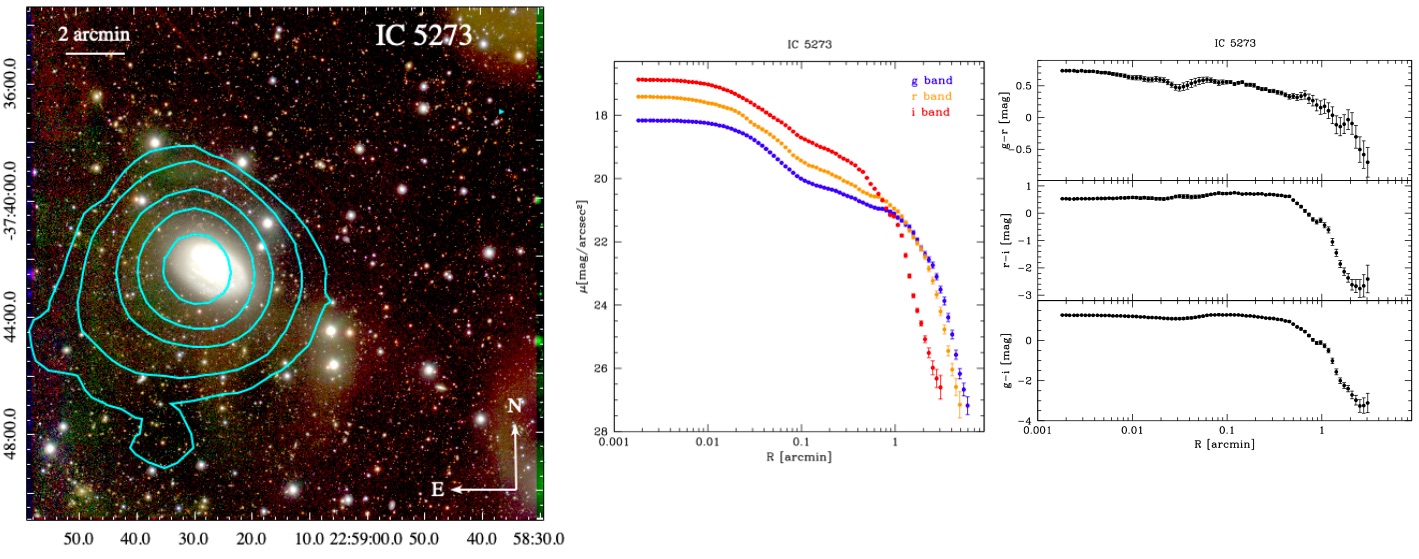}
	\caption{Same as Figure~\ref{fig:ic5270sp} but for IC~5273. The HI contour levels are $1$, $2.5$, $7.5$, $12.5$, and $20 \times 10^{19}$~cm$^{-2}$.}
	\label{fig:ic5273sp}
\end{figure*}

For each galaxy of the IC~1459 group, we report the results of the surface photometry, from Figure~\ref{fig:ic5270sp} to Figure~\ref{fig:ic5273sp}. The colour composite image (red channel for {\it i} band, green channel for {\it r} band, and blue channel for {\it g} band; left panel) extracted from the VST mosaic around the galaxy, with the HI map from the KAT$-7$ observations (cyan contours).
%; the white arrow indicates the direction of the BGG, IC~1459. 
We also report the azimuthally-averaged surface-brightness radial profile plotted in logarithmic scale as a function of the semi-major axis (middle panel), and the azimuthally-averaged extinction-corrected $g-r$, $r-i$, and $g-i$ colour profile as a function of the logarithmic semi-major axis (right panel). These plots are derived by the isophote fit from {\it g} (blue dots), {\it r} band (orange dots), and {\it i} band (red dots) VST images.

%---------------------------------------------- Figure X------------------------------------
%---------------------------------------------- end Figure X------------------------------------
%---------------------------------------------- Figure X------------------------------------
%---------------------------------------------- end Figure X------------------------------------
%---------------------------------------------- Figure X------------------------------------
%---------------------------------------------- end Figure X------------------------------------
%---------------------------------------------- Figure X------------------------------------
%---------------------------------------------- end Figure X------------------------------------
%---------------------------------------------- Figure X------------------------------------
%---------------------------------------------- end Figure X------------------------------------
%---------------------------------------------- Figure X------------------------------------
%---------------------------------------------- end Figure X------------------------------------
%---------------------------------------------- Figure X------------------------------------
%---------------------------------------------- end Figure X------------------------------------
%---------------------------------------------- Figure X------------------------------------
%---------------------------------------------- end Figure X------------------------------------

\end{document}